# INVESTIGATION OF THE DYNAMICS OF TRANSVERSE OSCILLATIONS OF A VERTICAL ROD UNDER GRAVITY, FRICTION, AND THERMAL EXPANSION

**E.E. Perepelkin[a,b,c,d,*], B.I. Sadovnikov[a], N.G. Inozemtseva[b,c], M.V. Klimenko[a]**

*[a] Faculty of Physics, Lomonosov Moscow State University, Moscow, 119991 Russia*
*[b] Moscow Technical University of Communications and Informatics, Moscow, 123423 Russia*
*[c] Dubna State University, Moscow region, Dubna,141980 Russia*
*[d] Joint Institute for Nuclear Research, Moscow region, Dubna,141980 Russia*
*[*]corresponding author*

## Abstract

The paper considers the mathematical formulation of the problem of transverse oscillations of a vertical rod under gravity, friction and external pulse effect, leading to thermal expansion of the rod. The dynamics of the system under consideration corresponds to the behavior of a fuel element (FE) in a pulsed reactor and is related to the dynamic stability of the processes occurring in it.

The FE dynamics is described by inhomogeneous linear differential equation of the fourth order with non-constant coefficients. Initial boundary conditions are not smooth since they correspond to the instant heating of the part of the FE surface exposed to neutron pulse radiation. The general solution of the homogeneous linear equation can be found using the concept of generalized functions expressed as a series expansion in terms of coordinate eigenfunctions dependent on $p$ parameter. The $p$ parameter is related with the FE mass and at some certain values results in bifurcation points of the boundary value problem for eigenfunctions. The partial solution can be found in several ways: using the Fourier transform, the method of Green's function, and in terms of series expansion by eigenfunctions.

Eigenfunction expansion coefficients in an explicit form have been obtained and the numerical solution accuracy has been estimated for some important particular cases. The results of the obtained exact solutions appear to be very close to the results of the ANSYS numerical estimations. In the future the obtained exact solutions will be used as input data to simulate self-consistent system dynamics of more than 50 FE. Therefore, the advantage of the exact solution in practical use is that its calculation is much faster as compared with the finite element method done with ANSYS.

**Key words:** oscillation equation, thermal expansion, stress-strain state, bifurcation point, nuclear reactors.

## Introduction

The problem of the stress-strain state is of great practical importance in applied fields. The works of L. Euler [1] played a significant role in building the foundation of the mathematical apparatus for describing such physical systems. Over the past centuries, many practical problems have been solved, studies of the properties of existing and newly created materials have been carried out, software packages have been written that allow modeling the dynamics of coupled 3D systems, taking into account strain calculations, external thermal and electromagnetic effects [2-6, 22].

Despite the high potential of computing tools, at the initial stage of creating an installation or analyzing the properties of a real working one, it is still important to build a «zero approximation» in the form of the simplest analytical model of the system. Before entering data into the calculation program, it is necessary to have an analytical model that captures the essence of the dynamics of a physical system. Such a model makes it possible to estimate the order of the main parameters of the system and proceed to the iterative process of numerical simulation and optimization on the computing architecture [7]. The importance of the analytical model is not only in initiating the simulation process, but also in its verification. Most modern physical problems describe complex nonlinear systems without exact solutions. When performing numerical simulation of such a system, the question always arises about the accuracy of the result, about the correct choice of parameters of the numerical scheme. The presence of an exact solution of the model problem makes it possible to estimate the correctness of the parameters of the numerical scheme.

An example of such a problem is to model the shape dynamics of fuel elements (FE) in periodic pulsed (pulsating) reactors [8]. In such experimental nuclear reactors, power is generated using reactivity modulation at a frequency of 5 to 50 Hz, depending on the particular design. As a result, fuel elements are subject to periodic pulsed thermal effects. The presence of the elastic properties of the fuel element and the periodic uneven change in its temperature leads to the occurrence of transverse oscillations. Such behavior of a fuel element can significantly affect the reactivity and power parameters, which in turn can affect the stability of the reactor operation. Interest in this problem has noticeably increased in connection with the recently proposed concept of a new NEPTUNE pulsed reactor, which in its parameters surpasses the existing pulsed neutron sources and those under construction for research on extracted beams [9].

The purpose of this paper is to build an analytical model to describe the dynamic bending effect of the fuel elements in NEPTUNE/IBR-2 pulsed reactor. The obtained results will be input data for future estimation of the reactor stability under the influence of periodically alternating temperature of the fuel elements.

It should be noted that the presented paper is devoted not only to consideration of a specific type of fuel elements/rods, but also to generalization of the mathematical description that operates with a set of parameters and to investigation of possible behavior of the entire system. This approach is especially important when designing new installations, where it is necessary to see an overall picture of possible features of the system behavior.

The work has the following structure. Section 1 describes the formulation of the problem. A model of a fuel element in the form of a vertical rod with various types of fastening in its upper and lower parts is considered. The rod is elastic and experiences the action of three forces: friction due to being in a viscous medium, its own weight and the force of uneven thermal expansion. As a result, the formulation is reduced to an initial-boundary value problem for an inhomogeneous linear differential equation with variable coefficients in partial derivatives. The second section (§2) is devoted to the construction of an exact solution to the initial-boundary value problem and the analysis of its properties. The solution is constructed in a factorized form as an expansion in terms of the eigenfunctions of some operator $L_p$, where $p$ is a parameter related to the rod's own weight. The equation for coordinate eigenfunctions has a bifurcation point defined by parameter $p$, and leading to various stable/unstable types of solutions. The time part of the solution satisfies the inhomogeneous equation, the solution of which is sought in two ways: by the Green's function method and the integral Fourier transformation. We have discovered that to solve the FE dynamics problem it is preferable to use the method of Green's function which significantly reduces the calculation time, especially for the case of a rectangular function $\beta$ as compared with the Fourier transform method. Section 3 considers a particular form ($p = 0$) of coordinate eigenfunctions expressed in terms of the Krylov functions. Explicit analytical expressions are obtained for solving the problem and approximating a partial solution of an inhomogeneous equation. The limiting transition of the system to a static strain state when exposed to thermal expansion has been studied in detail. In Section 4, numerical simulation of the dynamics of transverse oscillations of the rod is carried out for specific physical parameters corresponding to the IBR-2 reactor fuel element [8]. Using the mathematical apparatus of the theory of generalized functions, solutions are constructed for initial-boundary value problems with various types (smooth and discontinuous) of the initial conditions of the Cauchy problem and types of external force. The use of generalized functions made it possible to discover the process of beating in the damped transverse oscillations of the rod, caused by its instantaneous heating. The case of periodic thermal pulses of different relative duration is considered. The influence of the relative duration of the thermal pulse on the nature of the oscillation process has been studied in detail. A number of analytical expressions have been obtained that make it possible to significantly reduce the time of numerical simulation of the system. A comparison of the analytical solution with the numerical one made with the use of ANSYS software is presented in the Discussion chapter. The Conclusion contains the main conclusions of the paper.

## §1 Statement of the problem

Fig. 1 shows a geometric model of a fuel element in the form of a cylindrical beam in a vertical position with a rigidly fixed upper end. The lower end is not fixed (see Fig. 1). The rod consists of a metal tube, which has inside pellets with nuclear fuel. Under the action of a temperature gradient on the surface of a metal tube (see Fig. 1), uneven thermal expansion occurs, leading to the rod deformation. As the rod material has elastic properties it generates mechanical oscillations of the fuel element. Within the framework of the model under consideration, we will focus on in plane transverse oscillations, which can have a major impact on the change in the reactivity and power of the reactor.

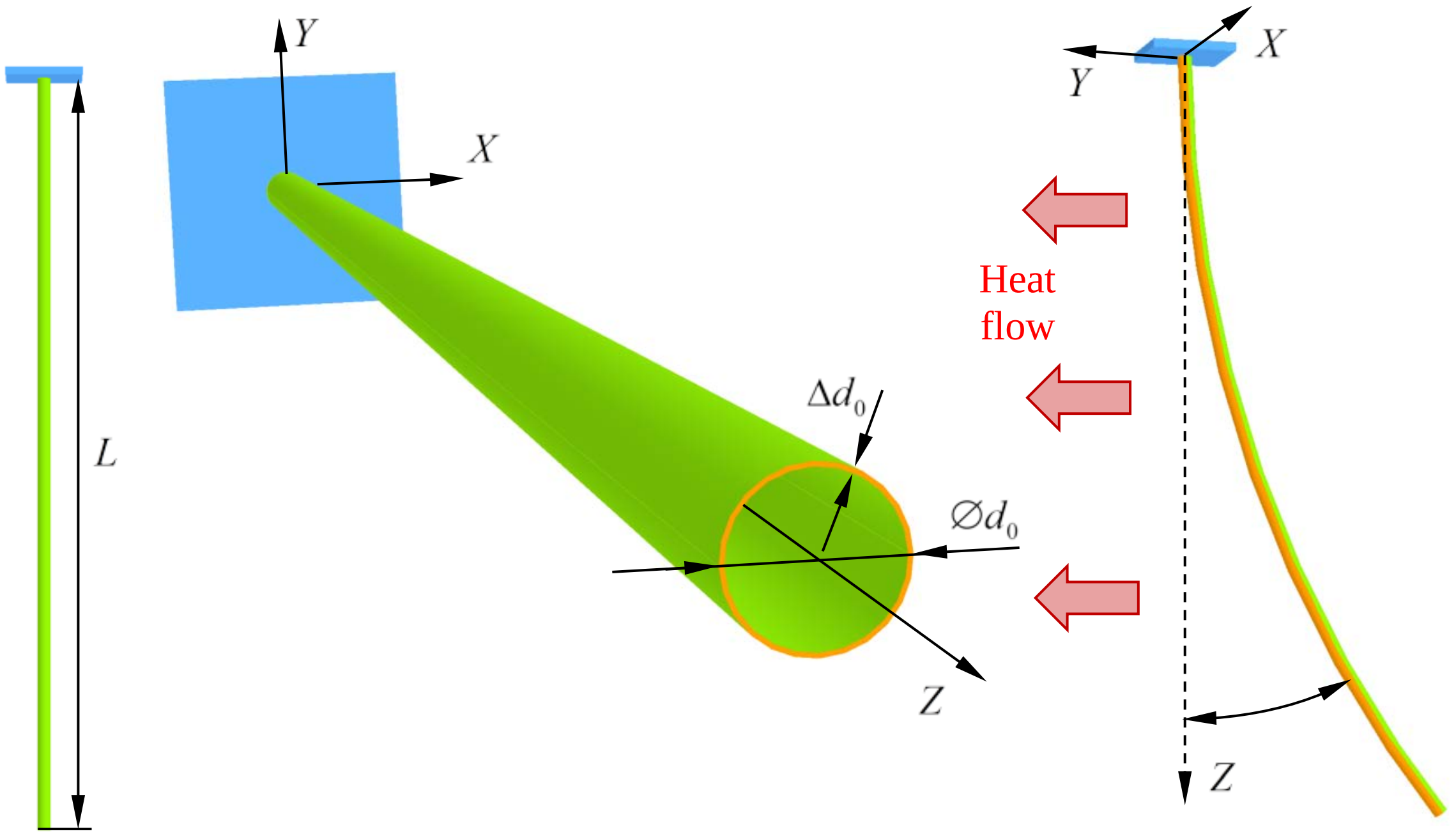

Fig. 1 Geometrical model of a fuel element.

Using the mathematical apparatus of the theory of strength of materials, we write the equation for the transverse oscillations of the beam [10, 11]. When constructing the oscillation equation, we will use the Huygens principle, which allows us to describe the static strain state, first and then move on to the dynamic representation by introducing inertial forces. Based on the symmetry of the model geometry, it is convenient to use a cylindrical coordinate system $(r, \varphi, z)$. Without loss of generality, we will assume that transverse oscillations occur in plane $YOZ$ (see Fig. 1). In terms of the theory of strength of materials, we will assume that axis $OZ$ coincides with the axis of the beam's «center of gravity», that is,

$$\int_S \vec{r} ds = 0, \quad \vec{r} = x\vec{e}_x + y\vec{e}_y, \tag{1.1}$$

where the integration is done over the cross section of the rod at point $z$ (see Fig. 2). Note that from a physical point of view, condition (1.1) corresponds to the position of the center of gravity of the section if the density of rod material $\rho$ satisfies the conditions:

$$\int_S \vec{r}\rho(r,\varphi,z)ds = \int_0^{2\pi} \vec{e}_r d\varphi \int_0^{R(\varphi)} \rho(r,\varphi,z) r^2 dr = 0,$$

$$\begin{cases} \int_0^{2\pi} \rho(r,\varphi,z)\cos\varphi d\varphi \int_0^{R(\varphi)} r^2 dr = 0, \\ \int_0^{2\pi} \rho(r,\varphi,z)\sin\varphi d\varphi \int_0^{R(\varphi)} r^2 dr = 0, \end{cases} \quad (1.2)$$

where function $R(\varphi)$ defines the boundary of the rod cross section. Conditions (1.2) can be satisfied, for example, with azimuthal density symmetry $\rho = \rho(r,z)$.

According to the Huygens principle, the differential relations of Zhuravsky [11] for a certain rod cross section (see Fig. 2) at some point in time $t$ will take the form:

$$\begin{cases} \frac{\partial \vec{M}(z,t)}{\partial z} = -\vec{m}(z,t) - Q_y(z,t)\vec{e}_x, \\ \frac{\partial \vec{Q}(z,t)}{\partial z} = -\vec{q}(z,t), \end{cases} \quad (1.3)$$

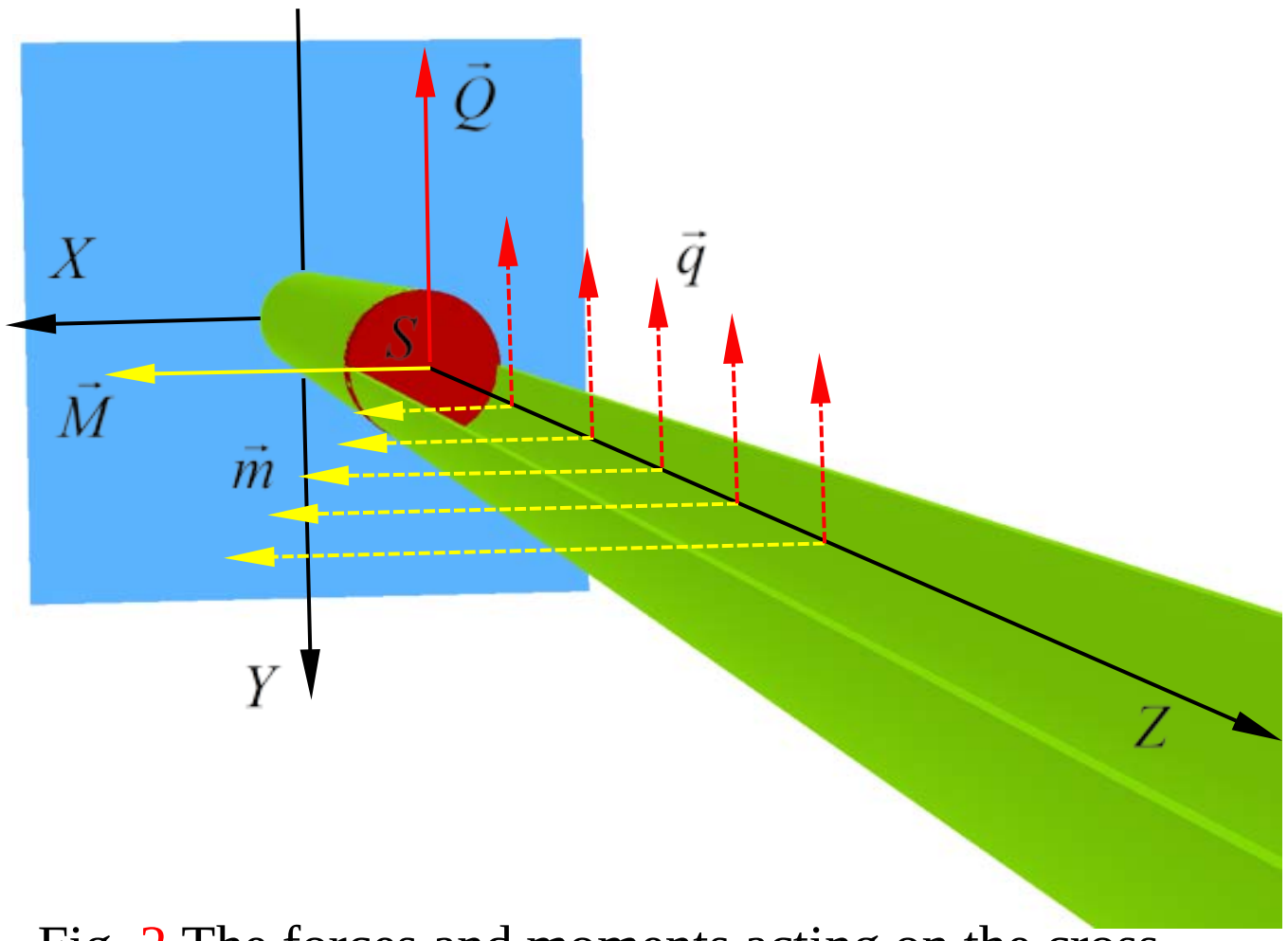

Fig. 2 The forces and moments acting on the cross section of the rod.

where $\vec{Q}(z,t) = \{Q_x, Q_y, Q_z\}$ is a force, and $\vec{M}(z,t)$ is the moment of the force, which affect the section; $\vec{q}(z,t)$ is a distributed force, and $\vec{m}(z,t)$ is the distributed moment; $\vec{e}_x, \vec{e}_y, \vec{e}_z$ are basis vectors. Since oscillations occur in plane $YOZ$, system (1.3) takes into account that $Q_x = 0$. The vector field of distributed moment $\vec{m}(z,t)$ has only $\vec{e}_x$ component, and the field of distributed force $\vec{q}(z,t)$ has $\vec{e}_y$ component.

Relations (1.3) are written for a certain cross section of the rod, which is affected (in the form of distributed moment $\vec{m}(z,t)$ and forces $\vec{q}(z,t)$) from its lower part (see Fig. 2). From system (1.3), the expression for distributed forces follows:

$$q_y(z,t) = \frac{\partial^2}{\partial z^2} M_x(z,t) + \frac{\partial}{\partial z} m_x(z,t), \quad (1.4)$$

which corresponds to the projection onto axis $OX$. Expression (1.4) can be derived from (1.3) by means of differentiation of the first equation over $z$ variable in the $\vec{e}_x$ projection.

Let function $u(z,t)$ determine the shift in the transverse direction (axis $OY$) from the equilibrium position of the rod axis (1.1) in point $z$ at point of time $t$. The moment of stiffness forces $M_x^{(D)}$ caused by the strain of the rod has the form [10, 11]:

$$M_x^{(D)}(z,t) = EJ_x \frac{\partial^2}{\partial z^2} u(z,t), \qquad (1.5)$$

where $E$ is a Young's modulus; $J_x$ is a geometrical moment of the section inertia of the rod about the axis parallel to axis $OX$ and passing through point $z$. Note that the geometrical moment of inertia differs from a usual moment of inertia by the absence of the density function $\rho$ under the integral, that is

$$J_x(z) = \int_{S(z)} y^2 ds, \qquad (1.6)$$

where integration in expression (1.6) is performed over cross section $S(z)$ of the rod, which in our case is constant. Thus, the value (1.5) has dimension $[N \cdot m]$.

The temperature of the opposite (on the $OY$ axis) walls of the rod $S_\pm = \{(r,\varphi,z): r = d_0/2, 0 \le \pm\varphi \le \pi\}$ (see Fig. 1) differs by $\pm\Delta T$ from the core temperature $T_0$. A linear approximation for the temperature distribution along the $OY$ axis will be considered in the framework of our model. The cross section of the rod has uneven heating, which is determined by the following relation:

$$T(y,z,t) = \iota \frac{y}{d_0} \langle T \rangle (z,t) = \iota \frac{y}{d_0} \tau(z) \beta(t), \qquad (1.7)$$

where $\iota$ is a dimensionless coefficient equal to the ratio of the temperature difference on the fuel element surfaces $S_\pm$ (see Fig. 1) to the average can temperature, which is calculated from the energy release gradient in the core; $\langle T \rangle (z,t)$ is the average temperature of the FE can, which is considered to be known from the numerical solution of the time dependent heat conduction equation. Since the solution of the heat conduction equation is sought in a factorized form $\langle T \rangle (z,t) = \tau(z)\beta(t)$, it can be represented (1.7) in terms of some functions $\tau(z)$ and $\beta(t)$. The graph of function $\tau(z)$ is given in Fig. 3 and can approximated by the function:

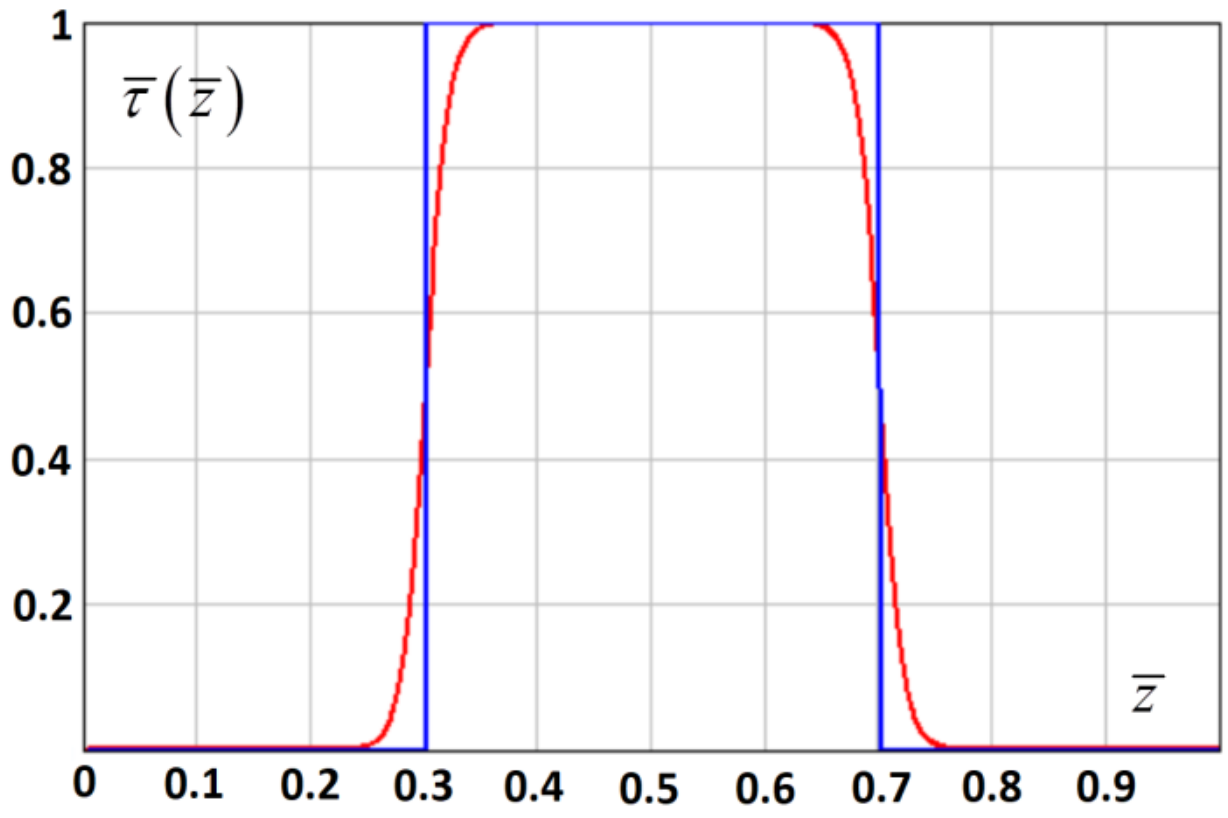

Fig. 3 Scale function of temperature distribution $\bar{\tau}(\bar{z})$ in the FE.

$$\tau(z) = \frac{1}{1+e^{\Upsilon(z_1 - z)}} + \frac{1}{1+e^{\Upsilon(z - z_2)}} - 1, \quad z_{1,2} = \frac{L}{2} \mp \Delta, \qquad (1.8)$$

where $\Delta = 0.2m$, $L = 1m$, $\Upsilon = 100/L$. We will also use the notation of function (1.8) with dimensionless variables $\bar{z}L = z$, $\Upsilon L = \bar{\Upsilon}$, $\tau(z) = \tau(\bar{z}L) = \bar{\tau}(\bar{z})$.

The choice of function (1.7) is due to the fact that the bending moment $M_x^{(T)}$ is only determined by the difference of temperatures on surfaces $S_+$ and $S_-$, although the heating of the FE can after the pulse causes an increase in temperature on both surfaces $S_\pm$.

According to the continuum mechanics the expressions for principal stresses (tensors) $\sigma_{xx}, \sigma_{yy}, \sigma_{zz}$ can be represented as:

$$\begin{aligned}
\sigma_{xx} &= \frac{E_x}{h}\left(1-\nu_{yz}^2\frac{E_z}{E_y}\right)(\varepsilon_x - \alpha_x\Delta T) + \frac{E_y}{h}\left(\nu_{xy} + \nu_{xz}\nu_{yz}\frac{E_z}{E_y}\right)(\varepsilon_y - \alpha_y\Delta T) + \\
&+ \frac{E_z}{h}(\nu_{xz} + \nu_{yz}\nu_{xy})(\varepsilon_z - \alpha_z\Delta T), \\
\sigma_{yy} &= \frac{E_y}{h}\left(\nu_{xy} + \nu_{xz}\nu_{yz}\frac{E_z}{E_y}\right)(\varepsilon_x - \alpha_x\Delta T) + \frac{E_y}{h}\left(1-\nu_{xz}^2\frac{E_z}{E_x}\right)(\varepsilon_y - \alpha_y\Delta T) + \\
&+ \frac{E_z}{h}\left(\nu_{yz} + \nu_{xz}\nu_{xy}\frac{E_y}{E_x}\right)(\varepsilon_z - \alpha_z\Delta T), \\
\sigma_{zz} &= \frac{E_z}{h}(\nu_{xz} + \nu_{yz}\nu_{xy})(\varepsilon_x - \alpha_x\Delta T) + \frac{E_z}{h}\left(\nu_{yz} + \nu_{xz}\nu_{xy}\frac{E_y}{E_x}\right)(\varepsilon_y - \alpha_y\Delta T) + \\
&+ \frac{E_z}{h}\left(1-\nu_{xy}^2\frac{E_y}{E_x}\right)(\varepsilon_z - \alpha_z\Delta T),
\end{aligned} \tag{1.9}$$

$$h = 1 - \nu_{xy}^2\frac{E_y}{E_x} - \nu_{yz}^2\frac{E_z}{E_y} - \nu_{xz}^2\frac{E_z}{E_x} - 2\nu_{xy}\nu_{yz}\nu_{xz}\frac{E_z}{E_x}, \tag{1.10}$$

where $\alpha$ is the temperature coefficient of linear expansion; $\nu$ stands for the Poisson coefficients; $\Delta T$ is the temperature difference; $\varepsilon$ stands for principal strain. Within the framework of the model under consideration the strains are assumed to be negligible ( $\varepsilon_x = \varepsilon_y = \varepsilon_z = 0$ ) and the transversal deformation has low impact on the longitudinal deformation ($\nu_{xz} = \nu_{yz} = 0$) and $E_x = E_y = E_z = E$ for low-amplitude oscillations. Another assumption is that the temperature deformation only essentially contributes to longitudinal direction ($\alpha_x = \alpha_y = 0,\ \alpha_z = \alpha_T$). Therefore, expressions (1.9) and (1.10) take the following form:

$$h = 1 - \nu_{xy}^2,$$

$$\sigma_{xx} = \sigma_{yy} = 0, \qquad \sigma_{zz} = -\alpha_T\Delta T\frac{E}{h}(1-\nu_{xy}^2) = -E\alpha_T\Delta T, \tag{1.11}$$

Taking expression (1.11) and Fig. 4 into account the distribution of longitudinal stresses inside the fuel element section, caused by thermal expansion can be represented as:

$$\sigma_{zz}(y,z,t) = E\alpha_T T(y,z,t). \tag{1.12}$$

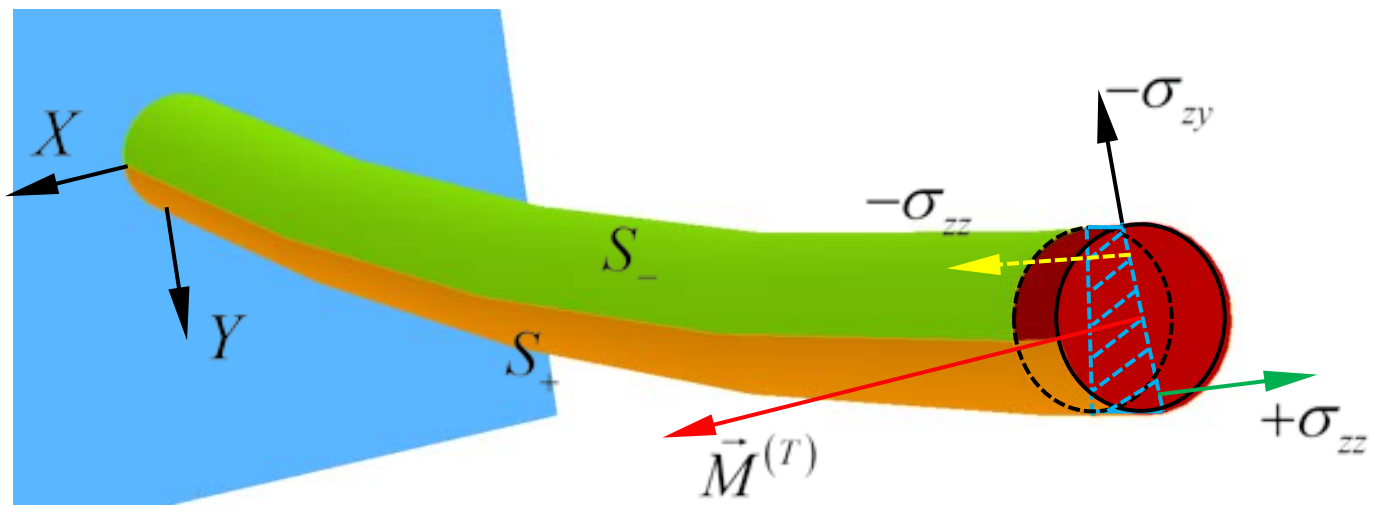

Fig. 4 Distribution of temperature stresses $\sigma_{zz}$, $\sigma_{zy}$

Due to thermal deformation (1.7), the longitudinal section of a cylindrical segment of length $dz$ changes from a rectangular shape to a trapezoid shape (blue hatching in Fig. 4). This shape of the longitudinal section characterizes the linear dependence of stress $\sigma_{zz}$ (1.12), (1.7).

Note that, in addition to longitudinal stresses (1.12), there are tangent stresses $\sigma_{zx}$ and $\sigma_{zy}$ in the fuel element cross section, which are due to thermal expansions in the transverse direction (see Fig. 4). Due to the symmetry of the problem, we will not take into account stresses $\sigma_{zx}$, and the contribution of stresses $\sigma_{zy}$ to the moment $M_x^{(T)}$ in the framework of the model under consideration is considered negligible.

Using expressions (1.6), (1.7), (1.12) force moment $M_x^{(T)}$ will take the form (see Fig. 4):

$$M_x^{(T)} = \int_S y\sigma_{zz} ds = \frac{\iota}{d_0} E\alpha_T \tau(z)\beta(t) \int_S y^2 ds = \frac{\iota}{d_0} EJ_x \alpha_T \tau(z)\beta(t),$$

$$M_x^{(T)}(z,t) = \frac{\alpha_T \iota}{d_0} EJ_x \langle T \rangle (z,t). \tag{1.13}$$

Distributed moment of forces $m_x(z,t)$ in expression (1.4) is determined by the action of the rod's own weight. Fig. 5 shows a model of an infinitely small rod element (red) with longitudinal dimension $dz$, which deviates from axis $OZ$ by a small amount $\Delta u(z,t)$. The remaining lower part of the rod pulls it down with a force (weight), $P_z(z) = \mathrm{m}(z) g$ where $g$ is the acceleration of gravity, and $\mathrm{m}(z)$ corresponds to the mass of the lower part of the rod:

$$P_z(z) = M_0 g\left(1 - \frac{z}{L}\right), \tag{1.14}$$

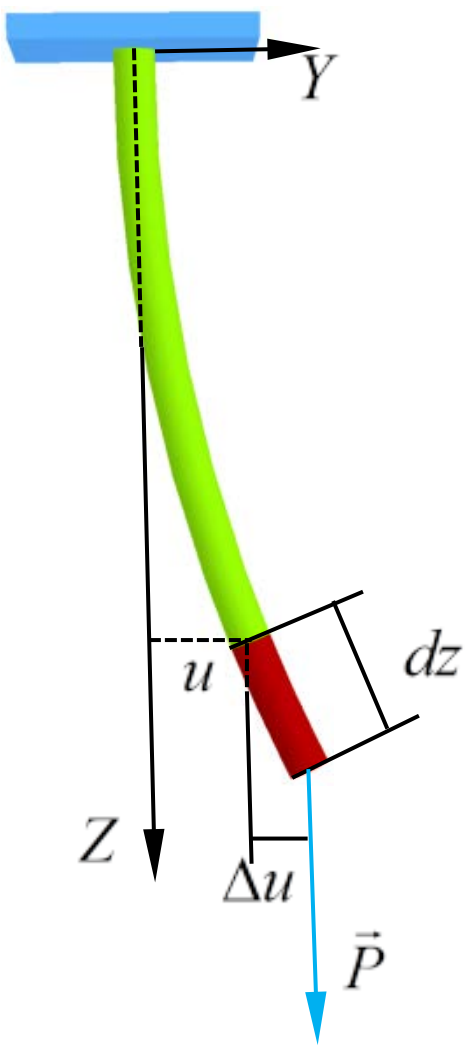

Fig. 5 Gravitation moment

where $M_0$ is the total FE mass. In expression (1.14) it is assumed that the density of FE mass is homogeneous in the longitudinal direction. As a result, distributed moment of forces $m_x(z,t)$ will take the form:

$$m_x(z,t)dz = P_z(z)\Delta u(z,t),$$

$$m_x(z,t) = P_z(z)\frac{\partial}{\partial z}u(z,t), \tag{1.15}$$

where the transition to the limit is made with respect to the increment $dz$.

Since the FE is in a viscous substance, in addition to the distributed force (1.4), we take into account the distributed force of viscous friction

$$f(z,t) = d_0 \eta \frac{\partial}{\partial t} u(z,t), \tag{1.16}$$

where $\eta$ is a friction coefficient. Substituting expressions (1.5), (1.13), (1.15) and (1.16) into expression for the distributed force (1.4) and equating it according to the Huygens principle to the force of inertia $-\rho S \frac{\partial^2}{\partial t^2} u(z,t)$, we obtain

$$\rho S \frac{\partial^2 u}{\partial t^2} + EJ_x \frac{\partial^4 u}{\partial z^4} + \frac{\partial}{\partial z}\left( P_z \frac{\partial u}{\partial z} \right) + d_0 \eta \frac{\partial u}{\partial t} + \frac{\alpha_T \iota}{d_0} EJ_x \beta \frac{\partial^2 \tau}{\partial z^2} = 0,$$

$$a_1 \frac{\partial^2 u}{\partial t^2} + a_2 \frac{\partial u}{\partial t} + \frac{\partial^4 u}{\partial z^4} + b \frac{\partial}{\partial z}\left( P_z(z) \frac{\partial u}{\partial z} \right) = \overline{f}(z,t), \tag{1.17}$$

where

$$a_1 = \frac{\rho S}{EJ_x}, \; a_2 = \frac{d_0 \eta}{EJ_x}, \; b = \frac{1}{EJ_x}, \; \overline{f}(z,t) = -\frac{\alpha_T \iota}{d_0} \beta(t) \frac{\partial^2 \tau(z)}{\partial z^2}.$$

It is worth to mention that equation (1.17) is actually an analog of the second Newton law for the rod cross-section $\rho S \frac{\partial^2 u}{\partial t^2} = -\left( f + q_y \right)$. Equation (1.17) must be supplemented with initial and boundary conditions:

$$u(z,0) = u_0(z), \qquad \left. \frac{\partial u}{\partial t} \right|_{t=0} = u_1(z), \tag{1.18}$$

$$\left. \frac{\partial u}{\partial t} \right|_{z=0} = \left. \frac{\partial u}{\partial z} \right|_{z=0} = u(0,t) = 0, \; \left. \frac{\partial^2 u}{\partial z^2} \right|_{z=L} = \left. \frac{\partial^3 u}{\partial z^3} \right|_{z=L} = 0.$$

Note that initial-boundary conditions (1.18) may have a different form depending on the way of fastening the FE. Boundary condition (1.18) corresponds to the rigid fastening of the rod in the origin of coordinates ($z = 0$) and to its free end at $z = L$. The derivatives $u'_z, u''_{zz}$ and $u'''_{zzz}$ under (1.18) condition determine the rotation angles, bending momenta and crosscut forces respectively. In the case of hinged fastening at $z = 0$ or $z = L$ the following condition is necessary

$$\left. \frac{\partial^2 u}{\partial z^2} \right|_{z=0,L} = \left. u \right|_{z=0,L} = 0. \tag{1.19}$$

Second initial condition (1.18) for $u_t(z,0)$ can be replaced with the condition

$$u\left(z, \overline{T}\right) = u_{\overline{T}}(z), \tag{1.20}$$

where $\overline{T}$ is a certain time interval, for example, equal to the pulse period for the function $\beta(t)$. Thus, various conditions for sets of initial-boundary conditions (1.18)-(1.20) are possible depending on the way of fastening the FE and the known information about its strain at some points in time.

As a result, we obtained the statement of problem (1.17)-(1.20).

## §2 Construction of the solution

Let us consider possible methods for finding the exact solution of problem (1.17)-(1.20). Equation (1.17) is a linear inhomogeneous partial differential equation, so its solution can be represented as a sum

$$u(z,t)=u_{g.h.}(z,t)+u_{p.i.}(z,t), \qquad (2.1)$$

where $u_{g.h.}(z,t)$ is the general solution of the homogeneous equation

$$a_1\frac{\partial^2 u}{\partial t^2}+a_2\frac{\partial u}{\partial t}+\frac{\partial^4 u}{\partial z^4}+b\frac{\partial}{\partial z}\left(P_z(z)\frac{\partial u}{\partial z}\right)=0, \qquad (2.2)$$

and $u_{p.i.}(z,t)$ is a partial solution of the inhomogeneous equation (1.17). Let us start with solving equation (2.2). We use the method of separation of variables, that is, we will seek solution (2.2) in a factorized form $u_{g.h.}(z,t)\sim \mathrm{Z}(z)\mathrm{T}(t)$, we get:

$$a_1\frac{\mathrm{T}''}{\mathrm{T}}+a_2\frac{\mathrm{T}'}{\mathrm{T}}=-\frac{\mathrm{Z}^{(4)}}{\mathrm{Z}}-\frac{b}{\mathrm{Z}}\left[P_z(z)\mathrm{Z}'\right]'=-\lambda=const, \qquad (2.3)$$

from here

$$\begin{cases} a_1\mathrm{T}''+a_2\mathrm{T}'+\lambda\mathrm{T}=0, \\ \mathrm{Z}^{(4)}+b\left[P_z(z)\mathrm{Z}'\right]'-\lambda\mathrm{Z}=0. \end{cases} \qquad (2.4)$$

When factoring, we will assume that function $\mathrm{Z}$ has the dimension of length (meters), and function $\mathrm{T}$ is dimensionless. Equation (2.4) for function $\mathrm{T}(t)$ has a solution:

at $\lambda\neq 0$:

$$\mathrm{T}(t)=e^{-\frac{a_2}{2a_1}t}\left[A\cos(\beta t)+B\sin(\beta t)\right], \qquad \beta=\frac{\sqrt{4a_1\lambda-a_2^2}}{2a_1}, \qquad (2.5)$$

at $\lambda=0$:

$$\mathrm{T}(t)=C+De^{-\frac{a_2}{a_1}t}, \qquad (2.6)$$

where in (2.5) the condition $4a_1\lambda>a_2^2$ is assumed to be satisfied. For $\lambda<0$ (the case of bifurcation considered below) and for $0<\lambda<\frac{a_2^2}{4a_1}$ function (2.5) will be exponential. Value $\frac{a_2}{2a_1}$ in the exponent (2.5) corresponds to the oscillation damping coefficient, and $A,B,C,D$ are constant values.

Let us construct a solution of the second equation from system (2.4) for the function $\mathrm{Z}(z)$ [13]. We write the equation in dimensionless form:

$$\frac{d^4\bar{\mathrm{Z}}}{d\bar{z}^4}+p\frac{d}{d\bar{z}}\left[(1-\bar{z})\frac{d\bar{\mathrm{Z}}}{d\bar{z}}\right]-\bar{\lambda}\bar{\mathrm{Z}}=0, \qquad (2.7)$$

where

$$\overline{z}L = z,\ Z(z) = Z(\overline{z}L) = \overline{Z}(\overline{z}),\ \ p = bL^2M_0g,\ \ \overline{\lambda} = \lambda L^4.$$

Taking into account the boundary conditions (1.17) for the function $\overline{Z}$, we obtain:

$$\overline{Z}(0) = \overline{Z}'(0) = 0, \qquad \overline{Z}''(1) = \overline{Z}'''(1) = 0. \tag{2.8}$$

For equation (2.7), we consider the case $\lambda = 0$, which corresponds to time independent equation (2.2). We define the function $\vartheta(\overline{z})$ as

$$\overline{Z}^{(\lambda=0)}(\overline{z}) = \overline{Z}^{(0)}(\overline{z}) = \int_0^{\overline{z}} \vartheta(x)\,dx, \tag{2.9}$$

then equation (2.7) becomes the equation for the function $\vartheta(\overline{z})$:

$$\vartheta''(\overline{z}) + p(1-\overline{z})\vartheta(\overline{z}) = 0, \tag{2.10}$$

$$\vartheta(0) = \vartheta'(1) = 0, \tag{2.11}$$

When changing variables $\overline{x} = (\overline{z}-1)\sqrt[3]{p}$, $\vartheta(\overline{z}) = \psi(\overline{x})$ equation (2.10) goes into the well-known Airy equation

$$\psi''(\overline{x}) - \overline{x}\psi(\overline{x}) = 0. \tag{2.12}$$

Airy equation (2.12) has a bifurcation point on the real axis, at which the form of the solution changes from «oscillatory» to «exponential». Two linearly independent solutions of equation (2.12) are Airy functions of the 1st $\mathrm{Ai}(\overline{x})$ and 2nd $\mathrm{Bi}(\overline{x})$ kind.

If parameter $p$ in equation (2.10) with boundary conditions (2.11) is arbitrary, then there is only a trivial solution $\vartheta(\overline{z}) \equiv 0$. Finding a non-trivial solution to boundary value problem (2.10)-(2.11) requires «sacrifice» free parameter $p$.

In fact, boundary value problem (2.10)-(2.11) is reduced to an eigenvalue problem with a weight function $(1-\overline{z})$, eigenfunctions $\{\vartheta_k(\overline{z})\}$ and eigenvalues $p_k$. The eigenfunctions $\{\vartheta_k(\overline{z})\}$ form an orthogonal basis:

$$\int_0^1 (1-\overline{z})\vartheta_k(\overline{z})\vartheta_n(\overline{z})\,d\overline{z} = 0,\ \ k \neq n. \tag{2.13}$$

Finding a non-trivial solution to boundary value problem (2.10)-(2.11) can be represented as a linear combination of functions $\mathrm{Ai}(\overline{x})$, $\mathrm{Bi}(\overline{x})$ or found by the Frobenius method through expansion in a power series:

$$\vartheta(\overline{z}, p) = \sum_{n=0}^{+\infty} \frac{(-p)^n}{c_n}(1-\overline{z})^{3n}, \tag{2.14}$$

$$c_0 = 1, \quad c_n = 3n(3n-1)c_{n-1}.$$

Taking into account boundary conditions (2.11) for functions (2.14), we obtain polynomial equation, which roots are eigenvalues $p_k$

$$\sum_{n=0}^{+\infty} \frac{(-1)^n}{c_n} p^n = 0. \tag{2.15}$$

Equation (2.15) has infinite set of roots $p = p_k$ that can be found numerically, for example, $p_1 \approx 7.84$, $p_2 \approx 56.0$, $p_3 \approx 148.5$, $p_4 \approx 285.4$. Eigenfunctions $\{\vartheta_k(\bar{z})\}$ are obtained from expression (2.14) as $\vartheta_k(\bar{z}) = \vartheta(\bar{z}, p_k)$. If series (2.14) admits term-by-term integration, then function (2.9) takes the form:

$$\bar{Z}_k^{(0)}(\bar{z}) = \bar{Z}^{(0)}(\bar{z}, p_k) = \sum_{n=0}^{+\infty} \frac{(-p_k)^n}{c_n} \frac{1-(1-\bar{z})^{3n+1}}{3n+1}. \tag{2.16}$$

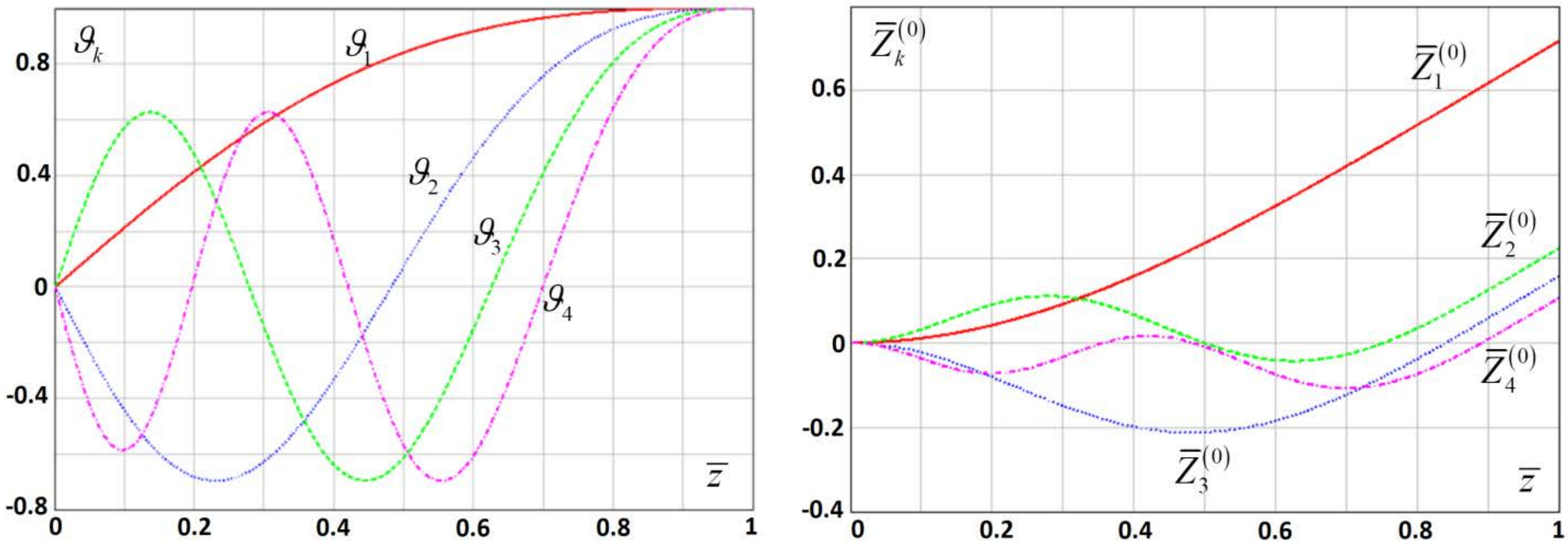

Fig. 6 Effect of bifurcation on eigenfunctions $\vartheta_k(\bar{z})$ and for $\bar{Z}_k^{(0)}(\bar{z})$.

As noted above, boundary value problem (2.9)-(2.10) has a bifurcation point. Indeed, when passing from the eigenvalue $p_1$ to $p_2$, a fundamental change occurs in the nature of the solution. Fig. 6 shows the graphs of functions $\vartheta_k(\bar{z})$ (left) and $\vartheta_k(\bar{z})$ (right) for eigenvalues $p_1,\ldots p_4$. It can be seen that for $p_1$ functions $\vartheta_1(\bar{z})$ and $\bar{Z}_1^{(0)}(\bar{z})$ are not periodic, but for $p_k$, $k>1$ a characteristic periodicity appears.

As a result, the original boundary value problem (2.7)-(2.8) for $\lambda = 0$ has a non-trivial solution, only for the values of $p = \frac{L^2 M_0 g}{EJ_x} = p_\ell$, $\ell \in \mathbb{N}$. The solution of the inhomogeneous equation (1.14) for $\lambda = 0$ and $p = p_\ell$ can be constructed if function is $\tau'' \sim \bar{Z}_\ell^{(0)}$, that is:

$$\frac{d^2\tau(z)}{dz^2} = \frac{\tau_0}{L^2} \bar{Z}_\ell^{(0)}(\bar{z}), \tag{2.17}$$

$$\tau(z,\ell)=\overline{\tau}(\overline{z},\ell)=\tau_0\int_0^{\overline{z}}dq\int_0^{q}\overline{Z}_\ell^{(0)}(s)\,ds+\tau_1\overline{z}+\tau_2,$$

$$\overline{\tau}(\overline{z},\ell)=\tau_0\sum_{n=0}^{+\infty}\frac{(-p_\ell)^n}{c_n(3n+1)}\left[\frac{\overline{z}^2}{2}-\frac{(1-\overline{z})^{3n+3}}{(3n+2)(3n+3)}\right]+\tau_1\overline{z}+\tau_2, \qquad (2.18)$$

where $\tau_0,\tau_1,\tau_2$ are some constants. Substituting (2.17) into equation (1.14) and representing its solution in the form $u_\ell(z,t)=\overline{Z}_\ell^{(0)}(\overline{z})\chi(t)$, we obtain

$$a_1\overline{Z}_\ell^{(0)}\chi''+a_2\overline{Z}_\ell^{(0)}\chi'=\overline{f}(z,t)=-\frac{\alpha_T\iota\tau_0}{d_0L^2}\beta(t)\overline{Z}_\ell^{(0)},$$

$$a_1\chi''+a_2\chi'=-\frac{\alpha_T\iota\tau_0}{d_0L^2}\beta(t). \qquad (2.19)$$

The solution of equation (2.19) can be found by the method of variation of an arbitrary constant. Introducing the notation $\kappa=\chi'$, we obtain

$$a_1\kappa'+a_2\kappa=-\frac{\alpha_T\iota\tau_0}{d_0L^2}\beta(t),\ \ \kappa(t)=-\frac{\alpha_T\iota\tau_0}{d_0L^2}\left(\frac{1}{a_1}\int_0^t\beta(\overline{t})e^{\frac{a_2}{a_1}\overline{t}}d\overline{t}+\kappa_0\right)e^{-\frac{a_2}{a_1}t}, \qquad (2.20)$$

where $\kappa_0$ is a constant value. Repeated integration of function (2.20) will give function $\chi(t)$. Solution $u_\ell(z,t)$ will satisfy boundary conditions (1.15) and the initial conditions:

$$u_0(z)=\overline{Z}_\ell^{(0)}(\overline{z})\chi(0), \qquad (2.21)$$
$$u_1(z)=\overline{Z}_\ell^{(0)}(\overline{z})\chi'(0).$$

In the case when $p\neq p_\ell$ at $\lambda=0$, it is possible to construct a solution of the inhomogeneous equation (1.14) by the Fourier method, expanding in terms of eigenfunctions $\vartheta_k(\overline{z})$:

$$u(z,t)=\sum_{k=1}^{+\infty}w_k(t)\overline{Z}_k^{(0)}(\overline{z}), \qquad (2.22)$$

where $w_k(t)$ are expansion coefficients that need to be determined. Substituting representation (2.22) into equation (1.17), we obtain

$$a_1\sum_{k=1}^{+\infty}\frac{d^2w_k}{dt^2}\overline{Z}_k^{(0)}+a_2\sum_{k=1}^{+\infty}\frac{dw_k}{dt}\overline{Z}_k^{(0)}+\frac{1}{L^4}\sum_{k=1}^{+\infty}w_k\frac{d^4\overline{Z}_k^{(0)}}{d\overline{z}^4}+\frac{b}{L^2}\frac{d}{d\overline{z}}\left(P_z(z)\sum_{k=1}^{+\infty}w_k\frac{d\overline{Z}_k^{(0)}}{d\overline{z}}\right)=\overline{f}(z,t),$$

$$a_1\sum_{k=1}^{+\infty}\frac{d^2w_k}{dt^2}\overline{Z}_k^{(0)}+a_2\sum_{k=1}^{+\infty}\frac{dw_k}{dt}\overline{Z}_k^{(0)}-\frac{1}{L^4}\sum_{k=1}^{+\infty}w_kp_k\frac{d}{d\overline{z}}\left[(1-\overline{z})\frac{d\overline{Z}_k^{(0)}}{d\overline{z}}\right]+$$

$$+\frac{bM_0g}{L^2}\frac{d}{d\overline{z}}\left[(1-\overline{z})\sum_{k=1}^{+\infty}w_k\frac{d\overline{Z}_k^{(0)}}{d\overline{z}}\right]=\overline{f}(z,t),$$

$$a_1\sum_{k=1}^{+\infty}\frac{d^2w_k}{dt^2}\bar{Z}_k^{(0)}+a_2\sum_{k=1}^{+\infty}\frac{dw_k}{dt}\bar{Z}_k^{(0)}+\frac{d}{d\bar{z}}(1-\bar{z})\sum_{k=1}^{+\infty}\frac{p-p_k}{L^4}w_k\frac{d\bar{Z}_k^{(0)}}{d\bar{z}}=\bar{f}(z,t), \qquad (2.23)$$

where according to (2.7) $\frac{d^4\bar{Z}_k^{(0)}}{d\bar{z}^4}=-p_k\frac{d}{d\bar{z}}\left[(1-\bar{z})\frac{d\bar{Z}_k^{(0)}}{d\bar{z}}\right]$. We multiply equation (2.23) by function $\bar{Z}_n^{(0)}$ and use the orthogonality property (2.13):

$$\begin{aligned}&a_1\sum_{k=1}^{+\infty}\frac{d^2w_k}{dt^2}\int_0^1\bar{Z}_n^{(0)}\bar{Z}_k^{(0)}d\bar{z}+a_2\sum_{k=1}^{+\infty}\frac{dw_k}{dt}\int_0^1\bar{Z}_n^{(0)}\bar{Z}_k^{(0)}d\bar{z}+\\&+\sum_{k=1}^{+\infty}\frac{p-p_k}{L^4}w_k\int_0^1\bar{Z}_n^{(0)}\frac{d}{dz}\left[(1-\bar{z})\frac{d\bar{Z}_k^{(0)}}{d\bar{z}}\right]d\bar{z}=\int_0^1\bar{Z}_n^{(0)}\bar{f}(z,t)\,d\bar{z}.\end{aligned} \qquad (2.24)$$

Let us introduce the notation:

$$c_{nk}=\int_0^1\bar{Z}_n^{(0)}\bar{Z}_k^{(0)}d\bar{z}, \qquad \bar{f}_n=\int_0^1\bar{Z}_n^{(0)}\bar{f}(z,t)\,d\bar{z}, \qquad (2.25)$$

and take into account that

$$\int_0^1\bar{Z}_n^{(0)}\frac{d}{d\bar{z}}\left[(1-\bar{z})\frac{d\bar{Z}_k^{(0)}}{d\bar{z}}\right]d\bar{z}=\bar{Z}_n^{(0)}(1-\bar{z})\frac{d\bar{Z}_k^{(0)}}{d\bar{z}}\bigg|_0^1-\int_0^1(1-\bar{z})\frac{d\bar{Z}_k^{(0)}}{d\bar{z}}\frac{d\bar{Z}_n^{(0)}}{d\bar{z}}d\bar{z}=-\|\vartheta_n\|^2\delta_{n,k}. \qquad (2.26)$$

Substituting (2.25) and (2.26) into equation (2.24), we obtain:

$$\begin{aligned}&a_1\sum_{k=1}^{+\infty}\frac{d^2w_k}{dt^2}c_{nk}+a_2\sum_{k=1}^{+\infty}\frac{dw_k}{dt}c_{nk}-\|\vartheta_n\|^2\sum_{k=1}^{+\infty}\frac{p-p_k}{L^4}w_k\delta_{n,k}=\bar{f}_n,\\&\sum_{k=1}^{+\infty}c_{nk}\left(a_1\frac{d^2w_k}{dt^2}+a_2\frac{dw_k}{dt}\right)+\|\vartheta_n\|^2\frac{p_n-p}{L^4}w_n=\bar{f}_n,\end{aligned} \qquad (2.27)$$

The solution of the system of differential equations (2.27) gives the value of the coefficients $w_k(t)$ for expansion (2.22). If $\iota\beta(t)=1^\circ K$, then a partial solution of system (2.27) takes the form:

$$w_n^{p.i.}=\frac{L^4\bar{f}_n}{\|\vartheta_n\|^2(p_n-p)}=const, \qquad (2.28)$$

or

$$w_n^{p.i.}=-\frac{\alpha_T L^2\cdot 1^\circ K}{\|\vartheta_n\|^2 d_0(p_n-p)}\int_0^1\bar{\tau}\frac{d\vartheta_n}{d\bar{z}}d\bar{z},$$

where $\bar{\tau}(\bar{z})=\tau(z)$ and it is assumed that $\bar{\tau}(1)=\bar{\tau}'(1)=0$. Function (2.22) satisfies boundary conditions (1.14), and the initial distribution must admit expansion into the series

$$u_0(z)=\sum_{k=1}^{+\infty} w_k(0)\bar{Z}_k^{(0)}(\bar{z}). \tag{2.29}$$

Let us pass to the consideration of the case $\lambda \neq 0$ for equation (2.7). The eigenvalue problem operator $L_p$ will depend on parameter $p$:

$$L_p\Phi_k(\bar{z},p)=\bar{\lambda}_k(p)\Phi_k(\bar{z},p), \tag{2.30}$$

$$\Phi_k(0,p)=\Phi_k'(0,p)=0, \qquad \Phi_k''(1,p)=\Phi_k'''(1,p)=0, \tag{2.31}$$

$$L_p=\frac{d^4}{d\bar{z}^4}+p\frac{d}{d\bar{z}}\left[(1-\bar{z})\frac{d}{d\bar{z}}\right],$$

where eigenfunctions $\Phi_k(\bar{z},p)$ and spectrum of eigenvalues $\bar{\lambda}_k(p)$ depend on the parameter $p$. Let us construct eigenfunctions $\Phi_k(\bar{z},p)$ using the Frobenius method

$$\Phi_k(\bar{z},p)=\sum_n \varepsilon_n(p,\bar{\lambda}_k)(1-\bar{z})^n, \tag{2.32}$$

where $\varepsilon_n(p,\bar{\lambda}_k)$ are the desired expansion coefficients. We substitute representation (2.32) into equation (2.30) and equate the coefficients at the same powers $(1-\bar{z})^n$, we obtain:

$$\varepsilon_n(p,\lambda)=\frac{(n-4)!}{n!}\left[\lambda\varepsilon_{n-4}-p\varepsilon_{n-3}(n-3)^2\right], \; n>3, \tag{2.33}$$

where coefficients $\varepsilon_0,\varepsilon_1,\varepsilon_2.\varepsilon_3$ are arbitrary values corresponding to the fourth order of differential equation (2.30). Since at $\bar{z}=1$ the beam is in a free state, then in the general case $\Phi_k(1,p)\neq 0$, therefore, $\varepsilon_0\neq 0$. Functions (2.32) are defined up to a normalization factor, so we set $\varepsilon_0=1$. Note that coefficients $\varepsilon_n$ have the dimension of length. From boundary conditions (2.31) it follows that:

$$\Phi_k(0,p)=0=\sum_{n=0}\varepsilon_n(p,\bar{\lambda}_k), \; \Phi'(0,p)=0=-\sum_{n=1}\varepsilon_n(p,\bar{\lambda}_k)n, \tag{2.34}$$

$$\Phi_k''(1,p)=0=\sum_{n=2}\varepsilon_n n(n-1)(1-1)^{n-2}=\varepsilon_2 2 \Rightarrow \varepsilon_2=0, \tag{2.35}$$

$$\Phi_k'''(1,p)=0=-\sum_{n=3}\varepsilon_n n(n-1)(n-2)(1-\bar{z})^{n-3}=-\varepsilon_3 3! \Rightarrow \varepsilon_3=0. \tag{2.36}$$

Expressions (2.34) are a system of equations requiring resolution with respect to eigenvalue $\bar{\lambda}_k$ and coefficient $\varepsilon_1$ of series (2.32). With respect to coefficient $\varepsilon_1$, system (2.34) is linear, and eigenvalues $\bar{\lambda}_k$ are the roots of a polynomial equation of infinite order:

$$\varepsilon_1=-\frac{1+\sum_{n=1}^{+\infty}\xi_n(\bar{\lambda})}{1+\sum_{s=1}^{+\infty}\zeta_s(\bar{\lambda})}, \qquad \frac{1+\sum_{n=1}^{+\infty}\xi_n(\bar{\lambda})}{1+\sum_{s=1}^{+\infty}\zeta_s(\bar{\lambda})}=\frac{1+\sum_{n=1}^{+\infty}n\xi_n(\bar{\lambda})}{1+\sum_{s=1}^{+\infty}s\zeta_s(\bar{\lambda})}, \tag{2.37}$$

where

$$\sum_{n=1}^{+\infty}\xi_n\left(\bar{\lambda}\right)=\frac{\bar{\lambda}}{4!}+\frac{\bar{\lambda}^2}{8!}+\frac{\bar{\lambda}^3}{12!}-4\frac{\bar{\lambda}p}{7!}+28\frac{\bar{\lambda}p^2}{10!}-12\frac{p\bar{\lambda}^2}{11!}-280\frac{p^3\bar{\lambda}}{13!}+...,$$

$$\sum_{s=1}^{+\infty}\zeta_s\left(\bar{\lambda}\right)=\frac{\bar{\lambda}}{5!}+\frac{\bar{\lambda}^2}{9!}+\frac{\bar{\lambda}^3}{13!}-\frac{p}{4!}+4\frac{p^2}{7!}-28\frac{p^3}{10!}+280\frac{p^4}{13!}-6\frac{\bar{\lambda}p}{8!}+52\frac{\bar{\lambda}p^2}{11!}-15\frac{\bar{\lambda}^2p}{12!}+...$$

Expressions (2.37) can be derived when substituting (2.33) in one of the (2.34) equations. The solution of system (2.37) can be found by numerical methods. First, root $\bar{\lambda}_k$ is found, and then value $\varepsilon_1^{(k)}$ (2.37) corresponding to it. Depending on the values of the parameter $p$, the spectrum of eigenvalues $\bar{\lambda}_k\left(p\right)$ may contain negative values, which are associated with unstable states (the presence of a bifurcation point). The spectrum of eigenvalues $\bar{\lambda}_k\left(p\right)$ contains an infinite set of positive values and a finite set of negative values.

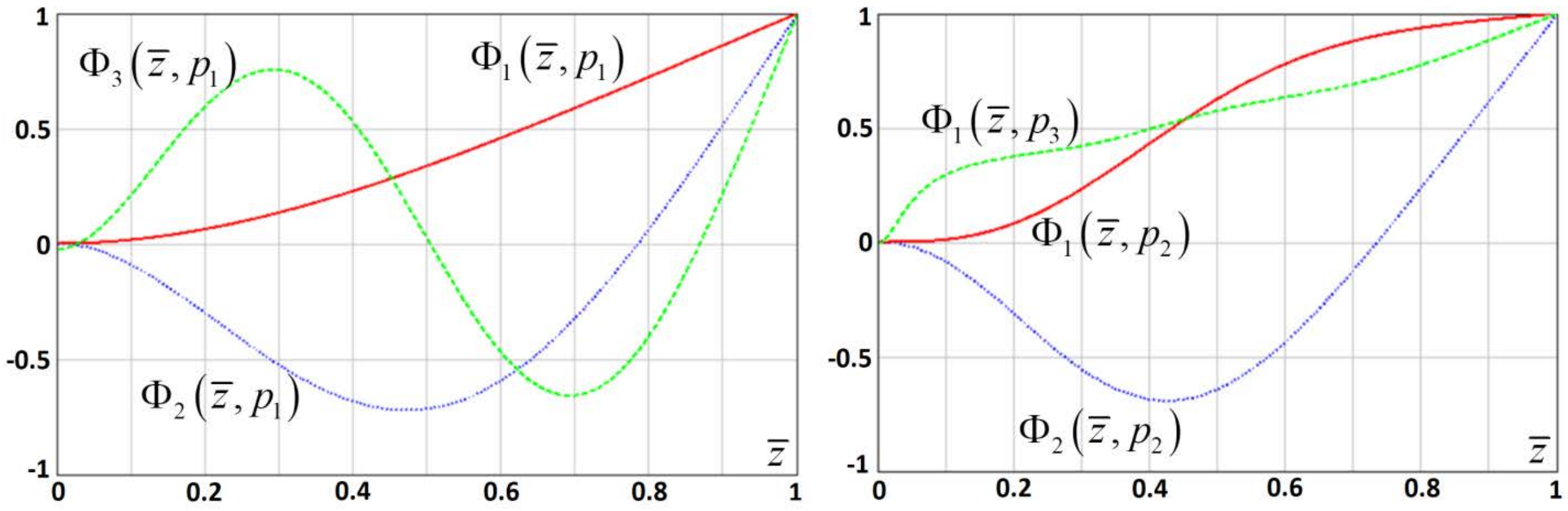

Fig. 7 Eigenfunctions $\Phi_k\left(\bar{z},p\right)$ for positive and negative eigenvalues $\bar{\lambda}_k$.

Fig. 7 (left) shows the graphs of the first three eigenfunctions (2.32) for $p_1=1$, with positive eigenvalues $\bar{\lambda}_k\left(p_1\right)$: $\bar{\lambda}_1\left(p_1\right)\approx 10.791$, $\bar{\lambda}_2\left(p_1\right)\approx 476.87$, $\bar{\lambda}_3\left(p_1\right)\approx 3781.581$ and coefficients $\varepsilon_1^{(k)}\left(p_1\right)$: $\varepsilon_1^{(1)}\left(p_1\right)\approx -1.378$, $\varepsilon_1^{(2)}\left(p_1\right)\approx -4.781$, $\varepsilon_1^{(3)}\left(p_1\right)\approx -7.842$. On the right in Fig. 7 graphs of eigenfunctions (2.32) with parameters $p_2=100>\bar{\lambda}_1$ and $p_3=500>\bar{\lambda}_2$ are shown. Two eigenfunctions $\Phi_1\left(\bar{z},p_2\right)$, $\Phi_2\left(\bar{z},p_2\right)$ with negative eigenvalues $\bar{\lambda}_1\left(p_2\right)\approx -145.547$, $\bar{\lambda}_2\left(p_2\right)\approx -425.778$ and coefficient values $\varepsilon_1^{(1)}\left(p_2\right)\approx -0.278$, $\varepsilon_1^{(2)}\left(p_2\right)\approx -3.777$ correspond to parameter $p_2$. One eigenfunction $\Phi_1\left(\bar{z},p_3\right)$ with a negative eigenvalue $\bar{\lambda}_1\left(p_3\right)\approx -460.90$ and coefficient $\varepsilon_1^{(1)}\left(p_3\right)\approx -1.163$ corresponds to value $p_3$.

Eigenfunctions (2.32) satisfying problem (2.30)-(2.31) are orthogonal

$$\int_0^1\Phi_k\left(\bar{z},p\right)\Phi_n\left(\bar{z},p\right)d\bar{z}=\left\|\Phi_n\right\|^2\delta_{n,k},\tag{2.38}$$

which simplifies the derivation of the equation for time coefficients $\bar{w}_k\left(t\right)$ in the expansion of the solution of inhomogeneous equation (1.16)

$$u(z,t)=\sum_{k=1}^{+\infty}\overline{w}_k(t)\Phi_k(\overline{z}). \tag{2.39}$$

Note that function (2.39) by virtue of conditions (2.31) satisfies boundary conditions (1.18) for equation (1.16). Substituting representation (2.39) into equation (1.17) and using properties (2.38), (2.30), we obtain:

$$a_1\sum_{k=1}^{+\infty}\overline{w}_k''\Phi_k + a_2\sum_{k=1}^{+\infty}\overline{w}_k'\Phi_k + \frac{1}{L^4}\sum_{k=1}^{+\infty}\overline{w}_k\left\{\frac{d^4\Phi_k}{d\overline{z}^4} + L^2 bM_0 g\frac{d}{d\overline{z}}\left[(1-\overline{z})\frac{d\Phi_k}{d\overline{z}}\right]\right\} = \overline{f}(z,t),$$

$$\sum_{k=1}^{+\infty}\left(a_1\overline{w}_k'' + a_2\overline{w}_k' + \frac{\overline{\lambda}_k}{L^4}\overline{w}_k\right)\Phi_k = \overline{f}(z,t), \tag{2.40}$$

$$a_1\overline{w}_n'' + a_2\overline{w}_n' + \frac{\overline{\lambda}_n}{L^4}\overline{w}_n = \overline{f}_n(t,p), \tag{2.41}$$

where

$$\overline{f}_n(t,p) = \frac{1}{\|\Phi_n\|^2}\int_0^1 \overline{f}(z,t)\Phi_n(\overline{z},p)\,d\overline{z}. \tag{2.42}$$

The resulting equation (2.41) is an inhomogeneous ordinary differential equation with constant coefficients. The right-hand side of (2.42), according to (1.17), can be represented as:

$$\overline{f}_n(t,p) = -\frac{\alpha_T t\overline{\tau}_n''}{d_0 L^2}\beta(t), \tag{2.43}$$

$$\overline{\tau}_n''(p) = \frac{1}{\|\Phi_n\|^2}\int_0^1 \overline{\tau}''(\overline{z})\Phi_n(\overline{z},p)\,d\overline{z}.$$

If function $\overline{\tau}(\overline{z})$ has property $\overline{\tau}(1)=\overline{\tau}'(1)=0$, then expression (2.43) can be rewritten in the form:

$$\overline{\tau}_n''(p) = \frac{1}{\|\Phi_n\|^2}\int_0^1 \overline{\tau}(\overline{z})\Phi_n''(\overline{z},p)\,d\overline{z}.$$

The uniqueness of the solution of equation (2.41) is determined by two initial conditions (1.18) or (1.20):

$$u(z,0)=\sum_{k=1}^{+\infty}\overline{w}_k(0)\Phi_k(\overline{z}) = u_0(z) = \sum_{k=1}^{+\infty}u_k^{(0)}\Phi_k(\overline{z}),$$

$$u_t(z,0)=\sum_{k=1}^{+\infty}\overline{w}_k'(0)\Phi_k(\overline{z}) = u_1(z) = \sum_{k=1}^{+\infty}u_k^{(1)}\Phi_k(\overline{z}),$$

$$\overline{w}_n(0) = u_n^{(0)} = \frac{1}{\|\Phi_n\|^2}\int_0^1 u_0(z)\Phi_n(\overline{z})\,d\overline{z}, \tag{2.44}$$

$$\overline{w}_n'(0) = u_n^{(1)} = \frac{1}{\|\Phi_n\|^2}\int_0^1 u_1(z)\Phi_n(\overline{z})\,d\overline{z}, \tag{2.45}$$

$$\overline{w}_n\left(\overline{T}\right)=u_n^{(T)}=\frac{1}{\left\|\Phi_n\right\|^2}\int_0^1 u_T\left(z\right)\Phi_n\left(\overline{z}\right)d\overline{z}, \tag{2.46}$$

where it is assumed that $u_0\left(z\right)$ and $u_1\left(z\right)$ admit expansion in functions $\Phi_k\left(\overline{z}\right)$. Conditions (2.44), (2.45) correspond to case (1.17), and conditions (2.44), (2.46) correspond to case (1.20). In both combinations (2.44), (2.45) and (2.44), (2.46), required function $\overline{w}_k\left(t\right)$ can be replaced by a function with zero conditions:

$$\overline{w}_n\left(t\right)=\tilde{w}_n\left(t\right)+u_n^{(1)}t+u_n^{(0)} \;\Rightarrow\; \tilde{w}_n\left(0\right)=\tilde{w}_n'\left(0\right)=0, \tag{2.47}$$

$$\overline{w}_n\left(t\right)=\widehat{w}_n\left(t\right)+\left(u_n^{(\overline{T})}-u_n^{(0)}\right)\frac{t}{\overline{T}}+u_n^{(0)} \;\Rightarrow\; \widehat{w}_n\left(0\right)=\tilde{w}_n\left(\overline{T}\right)=0, \tag{2.48}$$

where the change (2.47) corresponds to conditions (2.44), (2.45), and (2.48) is valid for the case (2.44), (2.46). Functions $\tilde{w}_n\left(t\right)$ and $\widehat{w}_n\left(t\right)$ will satisfy equation (2.41), but with different right-hand sides:

$$\tilde{f}_n\left(t,p\right)=\overline{f}_n\left(t,p\right)-\frac{\overline{\lambda}_n u_n^{(1)}}{L^4}t-\frac{\overline{\lambda}_n}{L^4}u_n^{(0)}-a_2u_n^{(1)}, \tag{2.49}$$

$$\widehat{f}_n\left(t,p\right)=\overline{f}_n\left(t,p\right)-\frac{\overline{\lambda}_n}{\overline{T}L^4}\left(u_n^{(\overline{T})}-u_n^{(0)}\right)t-\frac{\overline{\lambda}_n}{L^4}u_n^{(0)}-\frac{a_2}{\overline{T}}\left(u_n^{(\overline{T})}-u_n^{(0)}\right).$$

In view of the foregoing, the further presentation of the construction of solution $\overline{w}_k\left(t\right)$ to equation (2.41) will be carried out with boundary conditions, for example, (2.48). In this case, we will not change the notation for function $\overline{w}_k\left(t\right)$ to $\widehat{w}_n\left(t\right)$. The solution of equation (2.41) can be obtained in various ways. Each way has advantages and disadvantages depending on its subsequent numerical implementation.

**Eigenfunction expansion method**

Let us rewrite equation (2.41) in operator form:

$$\mathfrak{L}_n\overline{w}_n=\overline{\rho}\,\overline{f}_n, \tag{2.50}$$

$$\mathfrak{L}_k\overset{\det}{=}\frac{d}{dt}\overline{p}\left(t\right)\frac{d}{dt}-\overline{q}_k\left(t\right),\quad \overline{p}\left(t\right)\overset{\det}{=}a_1\overline{\rho}\left(t\right),\; \overline{q}_k\left(t\right)\overset{\det}{=}-\lambda_k\overline{\rho}\left(t\right),$$

$$\overline{\rho}\left(t\right)\overset{\det}{=}e^{2\gamma t},\qquad \gamma\overset{\det}{=}\frac{a_2}{2a_1}.$$

Let us find a set of eigenfunctions $\psi_n^{(k)}$ and eigenvalues $\lambda_n^{(k)}$ for the operator $\mathfrak{L}_k$ by solving the Sturm-Liouville problem:

$$\begin{cases}\mathfrak{L}_k\psi_n^{(k)}=\lambda_n^{(k)}\overline{\rho}\psi_n^{(k)},\\ \psi_n^{(k)}\left(0\right)=\psi_n^{(k)}\left(\overline{T}\right)=0.\end{cases} \tag{2.51}$$

The solution of problem (2.51) has the form:

$$\psi_n^{(k)}(t)=\sqrt{\frac{2}{\bar{T}}}e^{-\gamma t}\sin(\mu_n t),\ \ \mu_n\overset{\det}{=}\frac{\pi n}{\bar{T}},\qquad \tilde{\lambda}_n^{(k)}=\lambda_k-a_1\left(\gamma^2+\mu_n^2\right),\tag{2.52}$$

where eigenvalues $\tilde{\lambda}_n^{(k)}<0$ at $\lambda_k<0$ (the case of bifurcation). If $\lambda_k>0$, then positive and negative values of $\tilde{\lambda}_n^{(k)}$ are possible. In this case, $\tilde{\lambda}_n^{(k)}>0$ will lead to oscillation solutions (2.52), and $\tilde{\lambda}_n^{(k)}<0$ to exponentially damping ones.

Eigenfunctions $\psi_n^{(k)}$ do not explicitly depend on index «$k$», that is, all operators $\mathfrak{L}_k$ have the same set of eigenfunctions. Therefore, in the future, for simplicity of notation, we will omit index «$k$», that is $\psi_n^{(k)}\equiv\psi_n$. Note that the system of eigenfunctions $\{\psi_n\}$ (2.52) is orthonormal with weight $\rho$ (2.50), that is,

$$(\psi_n,\psi_m)\overset{\det}{=}\int_0^{\bar{T}}\psi_n(t)\rho(t)\psi_m(t)\,dt=\delta_{n,m}.\tag{2.53}$$

We represent the solution of equation (2.50) as an expansion in the basis $\{\psi_n\}$:

$$\bar{w}_k(t)=\sum_n c_n^{(k)}\psi_n(t).\tag{2.54}$$

Note that function (2.54) satisfies initial conditions (2.48). Substituting expansion (2.54) into equation (2.50) and taking into account the property of eigenfunctions (2.53), we obtain

$$\sum_n c_n^{(k)}\mathfrak{L}_k\psi_n=\sum_n c_n^{(k)}\tilde{\lambda}_n^{(k)}\bar{\rho}\psi_n=\bar{\rho}\bar{f}_k,$$
$$\sum_n c_n^{(k)}\tilde{\lambda}_n^{(k)}(\psi_n,\psi_j)=\int_0^{\bar{T}}\bar{f}_k(t)\bar{\rho}(t)\psi_j(t)\,dt\overset{\det}{=}\bar{f}_j^{(k)},$$
$$c_j^{(k)}=\frac{\bar{f}_j^{(k)}}{\tilde{\lambda}_j^{(k)}}.\tag{2.55}$$

Substituting the known coefficients (2.55) into expansion (2.54), we obtain the required solution of equation (2.50) with initial conditions (2.48):

$$\bar{w}_k(t)=\sum_n\frac{\bar{f}_n^{(k)}}{\tilde{\lambda}_n^{(k)}}\psi_n(t).\tag{2.56}$$

Note that a similar result (2.56) is obtained directly using the Green's function:

$$G_k(t,t')=\sum_n\frac{\psi_n(t)\psi_n(t')}{\tilde{\lambda}_n^{(k)}},\qquad \mathfrak{L}_kG_k(t,t')=\delta(t-t'),$$
$$\bar{w}_k(t)=\int_0^{\bar{T}}G_k(t,t')\bar{\rho}(t')\bar{f}_k(t')\,dt',$$

where $\delta(t-t')$ is the Dirac delta function.

As a result, solution (2.39) takes the form:

$$u(z,t)=\sum_{n,k=1}^{+\infty}\frac{\overline{f}_n^{(k)}}{\lambda_n^{(k)}}\psi_n(t)\Phi_k(\overline{z}). \tag{2.57}$$

In a particular case $\iota\beta(t)=1^{\circ}K$, solution (2.57) can be represented in the form:

$$\overline{f}_n^{(k)}=-\frac{\overline{\alpha}_T\mu_n\overline{\tau}_k''}{d_0L^2\left(\gamma^2+\mu_n^2\right)}\sqrt{\frac{2}{\overline{T}}}\left[1-(-1)^n e^{\gamma\overline{T}}\right], \tag{2.58}$$

$$u(z,t)=-\frac{\overline{\alpha}_T}{d_0L^2}\sqrt{\frac{2}{\overline{T}}}\sum_{n=1}^{+\infty}\mu_n\frac{1-(-1)^n e^{\gamma\overline{T}}}{\lambda_n^{(k)}\left(\gamma^2+\mu_n^2\right)}\psi_n(t)\sum_{k=1}^{+\infty}\overline{\tau}_k''\Phi_k(\overline{z}), \tag{2.59}$$

where value $\alpha_T$ after multiplication by $1^{\circ}K$ is a dimensionless value $\overline{\alpha}_T$.

As mentioned above, solution (2.56) and, consequently, expression (2.57) are written in the notation of (2.48), (2.49). If function (2.57) satisfies the initial conditions $u(z,0)=u(z,\overline{T})=0$, then the reverse transition (2.48) from function $\widehat{w}_n(t)$ to function $w_n(t)$ is unnecessary. In the general case of conditions (1.18), (1.20), it is necessary to make the reverse substitution:

$$\overline{w}_n(t)=\sum_n\frac{\overline{f}_n^{(k)}}{\lambda_n^{(k)}}\psi_n(t)+\left(u_n^{(\overline{T})}-u_n^{(0)}\right)\frac{t}{\overline{T}}+u_n^{(0)}, \tag{2.60}$$

where $w_n(0)=u_n^{(0)},\ w_n(\overline{T})=u_n^{(\overline{T})}$, and, according to (2.49), coefficients $\overline{f}_n^{(k)}$ have the form:

$$\begin{aligned}\overline{f}_n^{(k)}=&\int_0^{\overline{T}}\overline{f}_k(t,p)\overline{\rho}(t)\psi_n(t)\,dt-\frac{\overline{\lambda}_n}{\overline{T}L^4}\left(u_k^{(\overline{T})}-u_k^{(0)}\right)\int_0^{\overline{T}}t\overline{\rho}(t)\psi_n(t)\,dt-\\&-\left[\frac{\overline{\lambda}_n}{L^4}u_k^{(0)}-\frac{a_2}{\overline{T}}\left(u_k^{(\overline{T})}-u_k^{(0)}\right)\right]\int_0^{\overline{T}}\overline{\rho}(t)\psi_n(t)\,dt.\end{aligned} \tag{2.61}$$

The transition for initial conditions (2.44), (2.45) and (2.47) is done in a similar way.

**Fourier Integral Transform Method**

We represent solution $\overline{w}_n(t)$ of equation (2.41) as the sum of the general solution of the homogeneous equation and the partial solution of the inhomogeneous equation:

$$\overline{w}_n(t)=\overline{w}_n^{g.h.}(t)+\overline{w}_n^{p.i.}(t). \tag{2.62}$$

Solution $\overline{w}_n^{g.h.}(t)$ coincides with function (2.5) of equation (2.4). Let us find a partial solution $\overline{w}_n^{p.i.}(t)$. We use the Fourier integral to represent the Dirac delta function $\delta(t-t')=\frac{1}{2\pi}\int_{-\infty}^{+\infty}e^{i\omega(t-t')}d\omega$ and function $\overline{w}_n^{p.i.}(t)$:

$$\bar{f}_n(t)=\int_{-\infty}^{+\infty}\bar{f}_n(t')\delta(t-t')dt'=\frac{1}{2\pi}\int_{-\infty}^{+\infty}d\omega\int_{-\infty}^{+\infty}\bar{f}_n(t')e^{i\omega(t-t')}dt',$$

$$\bar{w}_n^{p.i.}(t)=\frac{1}{2\pi}\int_{-\infty}^{+\infty}d\omega\int_{-\infty}^{+\infty}\hat{R}_n(\omega)\,\bar{f}_n(t')e^{i\omega(t-t')}dt', \qquad (2.63)$$

where $\hat{R}_n(\omega)\in\mathbb{C}$ is the required function, which is the Fourier image of some function $R_n(t)$, i.e.

$$\hat{R}_n(\omega)=\frac{1}{\sqrt{2\pi}}\int_{-\infty}^{+\infty}R_n(t')e^{-i\omega t'}dt'. \qquad (2.64)$$

Substituting (2.63) into equation (2.41), we obtain:

$$\int_{-\infty}^{+\infty}d\omega\int_{-\infty}^{+\infty}\bar{f}_n(t')e^{i\omega(t-t')}\left[-a_1\omega^2\hat{R}+i\omega a_2\hat{R}+\lambda_n\hat{R}-1\right]dt'=0,$$

$$\hat{R}_n(\omega)=-\frac{1}{a_1\omega^2-\lambda_n-i\omega a_2}=-\frac{a_1\omega^2-\lambda_n+i\omega a_2}{\left(a_1\omega^2-\lambda_n\right)^2+\omega^2a_2^2}. \qquad (2.65)$$

Taking into account expression (2.43), partial solution (2.63) will take the form:

$$\bar{w}_n^{p.i.}(t)=-\frac{\alpha_T\iota\bar{\tau}''_n}{d_0L^2}\frac{1}{2\pi}\int_{-\infty}^{+\infty}\beta(t')e^{-i\omega t'}dt'\int_{-\infty}^{+\infty}\hat{R}_n(\omega)e^{i\omega t}d\omega=-\frac{\alpha_T\iota\bar{\tau}''_n}{d_0L^2\sqrt{2\pi}}\int_{-\infty}^{+\infty}\hat{\beta}(\omega)\hat{R}_n(\omega)e^{i\omega t}d\omega,$$

$$\bar{w}_n^{p.i.}(t)=-\frac{\alpha_T\iota\bar{\tau}''_n}{2\pi d_0L^2}(\beta*R_n)(t), \qquad (2.66)$$

where

$$\hat{\beta}(\omega)=\frac{1}{\sqrt{2\pi}}\int_{-\infty}^{+\infty}\beta(t')e^{-i\omega t'}dt', \qquad (\beta*R_n)(t)=\int_{-\infty}^{+\infty}\beta(t')R_n(t-t')dt'.$$

Thus, partial solution $\bar{w}_n^{p.i.}(t)$ of equation (2.41) is represented through the convolution of functions (2.64) $\beta(t)$ and $R_n(t)$. The general solution of the inhomogeneous equation (2.62) according to (2.5) and (2.66) takes the form:

$$\bar{w}_n(t)=e^{-\gamma t}\left[A_n\cos(\beta_nt)+B_n\sin(\beta_nt)\right]-\frac{\alpha_T\iota\bar{\tau}''_n}{2\pi d_0L^2}(\beta*R_n)(t), \qquad (2.67)$$

where $\beta_n=\frac{\sqrt{4a_1\lambda_n-a_2^2}}{2a_1}$. Substituting (2.67) into expansion (2.39), we obtain the solution of original equation (2.13):

$$u(z,t)=\sum_{n=1}^{+\infty}e^{-\gamma t}\left[A_n\cos(\beta_nt)+B_n\sin(\beta_nt)\right]\Phi_n(\bar{z})-\frac{\alpha_T\iota}{2\pi d_0L^2}\sum_{n=1}^{+\infty}\bar{\tau}''_n(\beta*R_n)(t)\Phi_n(\bar{z}), \qquad (2.68)$$

where the free constants $A_n$ and $B_n$ are determined based on initial conditions (1.15):

$$\sum_{n=1}^{+\infty} u_n^{(0)}\Phi_n(\overline{z}) + \frac{\alpha_T l}{2\pi d_0 L^2}\sum_{n=1}^{+\infty}\overline{\tau}_n''(\beta * R_n)(0)\Phi_n(\overline{z}) = \sum_{n=1}^{+\infty} A_n\Phi_n(\overline{z}), \tag{2.69}$$

$$\sum_{n=1}^{+\infty} u_n^{(1)}\Phi_n(\overline{z}) + \frac{\alpha_T l}{2\pi d_0 L^2}\sum_{n=1}^{+\infty}\overline{\tau}_n''(\beta * R_n)'(0)\Phi_n(\overline{z}) = \sum_{n=1}^{+\infty}(\mathrm{\beta}_n B_n - \gamma A_n)\Phi_n(\overline{z}).$$

Given the orthogonality of functions $\Phi_n(\overline{z})$, expressions (2.69) will take the form:

$$A_k = u_k^{(0)} + \frac{\alpha_T l}{2\pi d_0 L^2}\overline{\tau}_k''(\beta * R_k)(0), \tag{2.70}$$

$$B_k = \frac{1}{\mathrm{\beta}_k}u_k^{(1)} + \frac{\alpha_T l}{2\pi \mathrm{\beta}_k d_0 L^2}\overline{\tau}_k''(\beta * R_k)'(0) + \frac{\gamma}{\mathrm{\beta}_k}A_k.$$

Substituting the value of coefficients (2.70) into expression (2.68), we obtain the final form of the solution of original equation (1.17).

Finding the general solution (2.68) requires calculating the sum over functions (2.5) and convolution (2.66) with the Fourier inversion for function $R_n(t)$. Note that the calculation of the convolution and Fourier transformation requires numerical integration over an infinite region. The general solution $\overline{w}_n(t)$ (2.56) is obtained by calculating the integral (2.55) over a bounded region and summing the series.

**§ 3 Special case of «light-weight rod»**

In a number of practical cases it is possible to neglect the contribution of the force of gravity $P_z(z)$ to the dynamics of the motion of the rod, then initial equation (1.17) is significantly simplified:

$$a_1\frac{\partial^2 u}{\partial t^2} + a_2\frac{\partial u}{\partial t} + \frac{\partial^4 u}{\partial z^4} = \overline{f}(z,t). \tag{3.1}$$

Since equation (3.1) is a particular case of original equation (1.17), the above constructions of the solution remain valid. Due to the «absence» of gravity ($p = 0$), the spectrum of eigenvalues $\overline{\lambda}_k(p)$ will be positive and fixed. The form (2.32) of the eigenfunctions $\Phi_k(\overline{z},0)$ can be obtained explicitly through the functions of A.N. Krylov $\mathrm{K}_j$, $j = 1...4$ [12]:

$$\mathrm{K}_1(z) = \frac{1}{2}(\mathrm{ch}\, z + \cos z),\quad \mathrm{K}_2(z) = \frac{1}{2}(\mathrm{sh}\, z + \sin z), \tag{3.2}$$

$$\mathrm{K}_3(z) = \frac{1}{2}(\mathrm{ch}\, z - \cos z),\quad \mathrm{K}_4(z) = \frac{1}{2}(\mathrm{sh}\, z - \sin z),$$

$$\mathrm{K}_1(z) = \mathrm{K}_2'(z) = \mathrm{K}_3''(z) = \mathrm{K}_4'''(z) = \mathrm{K}_1^{(4)}(z),\ \mathrm{K}_1'(z) = \mathrm{K}_4(z), \tag{3.3}$$

$$\mathrm{K}_1(0) = 1,\ \mathrm{K}_2(0) = \mathrm{K}_3(0) = \mathrm{K}_4(0) = 0,$$

where the function argument $\mathrm{K}_j$ is a dimensionless variable. A linear combination of functions (3.2) $W(\bar{z})$ satisfies equation (2.30) with $p=0$ and $\bar{\lambda}=\alpha^4$:

$$W(\bar{z})=C_1\,\mathrm{K}_1(\alpha\bar{z})+C_2\,\mathrm{K}_2(\alpha\bar{z})+C_3\,\mathrm{K}_3(\alpha\bar{z})+C_4\,\mathrm{K}_4(\alpha\bar{z}), \qquad (3.4)$$

where $C_j$ are constants having the dimension of length. Substituting function (3.4) into boundary conditions (2.31), taking into account the properties (3.3), gives the spectrum of positive eigenvalues $\bar{\lambda}_k=\alpha_k^4$ and the set of eigenfunctions $\Phi_k(\bar{z},0)\overset{\text{det}}{=}W_k(\bar{z})=W(\alpha_k\bar{z})$:

$$C_1=C_2=0, \quad C_3=2, \; C_4(\alpha_k)=-2\frac{\mathrm{K}_1(\alpha_k)}{\mathrm{K}_2(\alpha_k)}=-2\frac{\mathrm{K}_4(\alpha_k)}{\mathrm{K}_1(\alpha_k)}, \qquad (3.5)$$

$$\cos\alpha_k\,\mathrm{ch}\,\alpha_k=-1. \qquad (3.6)$$

The orthogonality condition for functions $W_k$ can be justified by the Betti theorem (on the reciprocity of works) [11] $(W_s,W_k)=\delta_{nk}$.

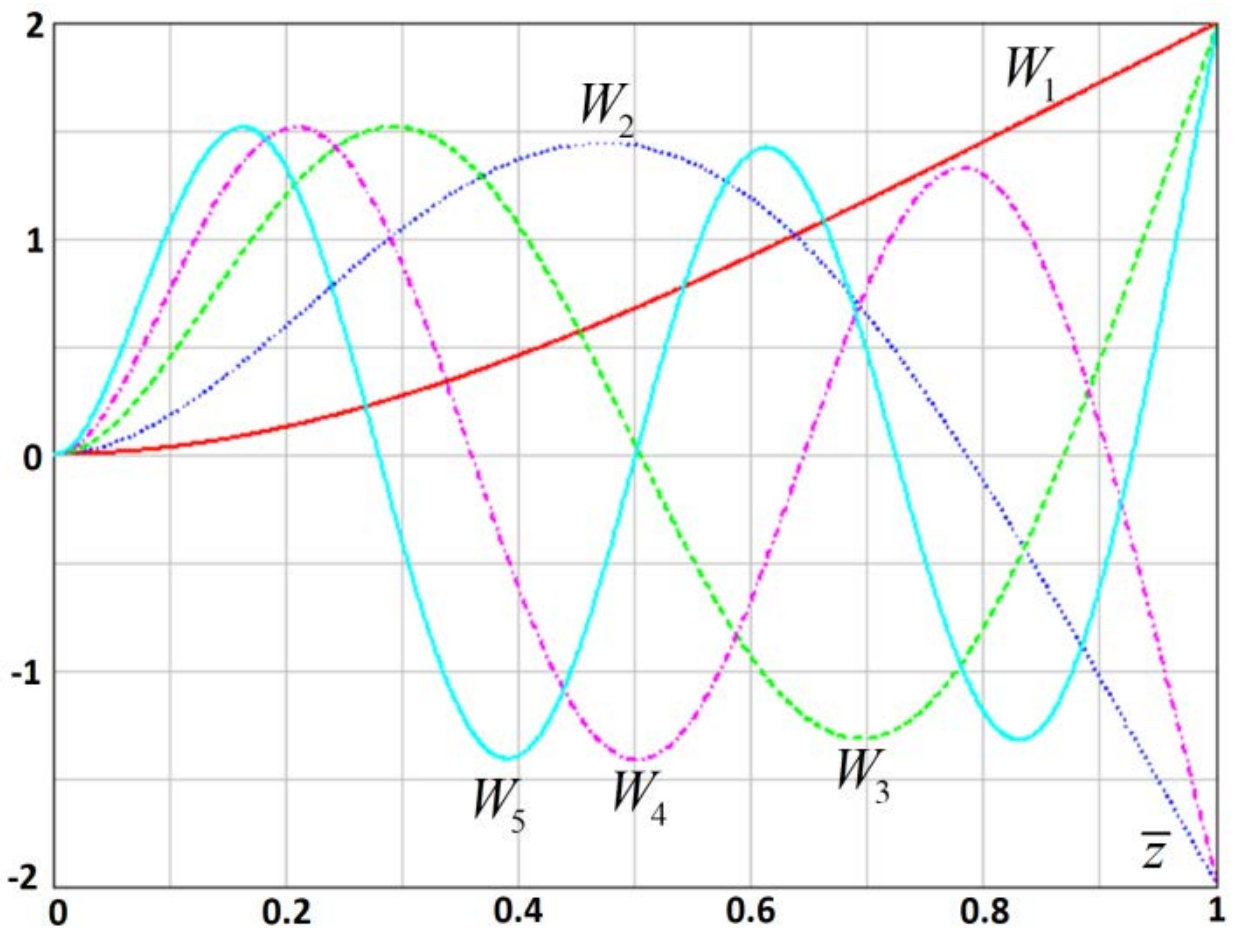

Fig. 8 Eigenfunctions $W_k$.

Fig. 8 shows the first five functions $W_k$. A comparison of Fig. 7 (left) and Fig. 8 shows a similar behavior for $W_1(\bar{z})$, $\Phi_1(\bar{z},p_1)$ and $W_3(\bar{z})$, $\Phi_3(\bar{z},p_1)$. Functions $W_2(\bar{z})$, $\Phi_2(\bar{z},p_1)$ are similar up to a sign. A significant difference between functions $W_k$ is observed from functions $\Phi_k$ shown in Fig. 7 (right). As shown in §2, functions $\Phi_1(\bar{z},p_2)$, $\Phi_1(\bar{z},p_3)$ and $\Phi_2(\bar{z},p_2)$, and unlike functions $W_k$ have negative eigenvalues.

For sufficiently large values $k\gg 1$, the approximation for eigenvalues $\bar{\lambda}_k=\alpha_k^4$ and eigenfunctions $W_k$ of the following form is valid:

$$\alpha_k\approx\pi(k-1/2), \qquad (3.7)$$

$$W_k(\bar{z})\approx e^{-\alpha_k\bar{z}}+\sin(\alpha_k\bar{z})-\cos(\alpha_k\bar{z})-e^{\alpha_k(\bar{z}-1)}\left[e^{-\alpha_k}+(-1)^k\right].$$

Expressions (3.7) can be effectively used in the numerical summation of series (see §2).

In §2, the Fourier integral transformation method for finding the function $\bar{w}_n^{p.i.}(t)$ is considered. Note that a partial solution $\bar{w}_n^{p.i.}(t)$ of equation (2.41) can be simplified for eigenfunctions $W_k$ in the case when $\iota\beta(t)=1^\circ K$. From expression (2.43) it follows that

$$\overline{f}_k(t) = -\frac{\alpha_T \iota \overline{\tau}''_k}{d_0 L^2}\beta(t) = -\frac{\overline{\alpha}_T \overline{\tau}''_k}{d_0 L^2} = const, \qquad (3.8)$$

where $\overline{\alpha}_T = \alpha_T \cdot 1^\circ K$ . From here

$$\overline{w}_\kappa^{p.i.}(t) = \frac{\overline{f}_k L^4}{\alpha_k^4} = -\frac{\overline{\alpha}_T L^2 \overline{\tau}''_k}{d_0 \alpha_k^4}. \qquad (3.9)$$

As a result, partial solution $u_{p.i.}(z,t)$ (2.1) of equation (1.14), taking into account (3.9), will not depend on time:

$$u_{p.i.}(z,t) = \sum_{n=1}^{+\infty} \overline{w}_n^{ч.н.}(t) W_n(\overline{z}) = -\frac{\overline{\alpha}_T L^2}{d_0} \sum_{n=1}^{+\infty} \frac{\overline{\tau}''_n}{\alpha_n^4} W_n(\overline{z}) = u_{p.i.}(z). \qquad (3.10)$$

Expression (3.10) can be simplified using the property of functions $W_k$ (2.30):

$$\frac{d^4}{d\overline{z}^4} u_{p.i.}(z) = -\frac{\overline{\alpha}_T L^2}{d_0} \sum_{n=1}^{+\infty} \overline{\tau}''_n W_n(\overline{z}) = -\frac{\overline{\alpha}_T L^2}{d_0} \overline{\tau}''(\overline{z}) = -\frac{\overline{\alpha}_T L^4}{d_0} \tau''(z), \qquad (3.11)$$

where (2.43) is taken into account, and function $\tau''(z)$ is known explicitly from the formulation of the problem (see §1, (1.8)). Integrating expression (3.11), we obtain:

$$\frac{d^2}{d\overline{z}^2} u_{p.i.}(z) = -\frac{\overline{\alpha}_T L^2}{d_0} \overline{\tau}(\overline{z}) + c_1 \overline{z} + c_2, \qquad (3.12)$$

or

$$u_{p.i.}(z) = -\frac{\overline{\alpha}_T L^2}{d_0} \int_0^{\overline{z}} d\overline{s} \int_0^{\overline{s}} \overline{\tau}(v)\, dv + c_1 \overline{z}^3 + c_2 \overline{z}^2 + c_3 \overline{z} + c_4,$$

where $c_1, c_2, c_3, c_4$ are some constants. Since a partial solution is sought, constants $c_j$, $j = 1...4$ can be taken equal to zero. Substituting the representation of function $\overline{\tau}$ in explicit form (1.8) into expression (3.12), we calculate the integral (3.12):

$$\int_0^{\overline{z}} d\overline{s} \int_0^{\overline{s}} \overline{\tau}(v)\, dv = I_{\overline{\Upsilon}}(\overline{z}, \overline{z}_1) + I_{-\overline{\Upsilon}}(\overline{z}, \overline{z}_2) - \frac{\overline{z}^2}{2}, \quad I_{\overline{\Upsilon}}(\overline{z}, \overline{z}_0) \overset{\det}{=} \int_0^{\overline{z}} d\overline{s} \int_0^{\overline{s}} \frac{dv}{1 + e^{\overline{\Upsilon}(\overline{z}_0 - v)}}, \qquad (3.13)$$

where the integration variables and $\overline{\Upsilon}$, $\overline{z}_1, \overline{z}_2$, are dimensionless values. It follows from expression (3.13) that it suffices to find only the integral $I_{\overline{\Upsilon}}(\overline{z}, \overline{z}_0)$:

$$\int_0^{\overline{s}} \frac{1}{1 + e^{\overline{\Upsilon}(\overline{z}_0 - v)}} dv = -\frac{1}{\overline{\Upsilon}} \int_{\overline{\Upsilon}\overline{z}_0}^{\overline{\Upsilon}(\overline{z}_0 - \overline{s})} \frac{1}{1 + e^x} dx = -\frac{1}{\overline{\Upsilon}} \int_{1+e^{\overline{\Upsilon}\overline{z}_0}}^{1+e^{\overline{\Upsilon}(\overline{z}_0 - \overline{s})}} \frac{dy}{y(y-1)} = -\frac{1}{\overline{\Upsilon}} \int_{1+e^{\overline{\Upsilon}\overline{z}_0}}^{1+e^{\overline{\Upsilon}(\overline{z}_0 - \overline{s})}} \frac{dy}{y-1} + \frac{1}{\overline{\Upsilon}} \int_{1+e^{\overline{\Upsilon}\overline{z}_0}}^{1+e^{\overline{\Upsilon}(\overline{z}_0 - \overline{s})}} \frac{dy}{y},$$

$$\int_0^{\overline{s}} \frac{1}{1 + e^{\overline{\Upsilon}(\overline{z}_0 - v)}} dv = \overline{s} + \frac{1}{\overline{\Upsilon}} \ln \frac{1 + e^{\overline{\Upsilon}(\overline{z}_0 - \overline{s})}}{1 + e^{\overline{\Upsilon}\overline{z}_0}},$$

$$I_{\bar{\Upsilon}}\left(\bar{z},\bar{z}_0\right)=\frac{\bar{z}^2}{2}-\frac{\bar{z}}{\bar{\Upsilon}}\ln\left(1+e^{\bar{\Upsilon}\bar{z}_0}\right)-\frac{1}{\bar{\Upsilon}^2}\int\limits_{\bar{\Upsilon}\bar{z}_0}^{\bar{\Upsilon}(\bar{z}_0-\bar{z})}\ln\left(1+e^x\right)dx=\frac{\bar{z}^2}{2}-\frac{\bar{z}}{\bar{\Upsilon}}\ln\left(1+e^{\bar{\Upsilon}\bar{z}_0}\right)+\frac{1}{\bar{\Upsilon}^2}\int\limits_{1+e^{\bar{\Upsilon}\bar{z}_0}}^{1+e^{\bar{\Upsilon}(\bar{z}_0-\bar{z})}}\frac{\ln y}{1-y}dy,$$

$$I_{\bar{\Upsilon}}\left(\bar{z},\bar{z}_0\right)=\frac{\bar{z}^2}{2}-\frac{\bar{z}}{\bar{\Upsilon}}\ln\left(1+e^{\bar{\Upsilon}\bar{z}_0}\right)-\frac{1}{\bar{\Upsilon}^2}\mathrm{Li}_2\left(-e^{\bar{\Upsilon}\bar{z}_0}\right)+\frac{1}{\bar{\Upsilon}^2}\mathrm{Li}_2\left(-e^{\bar{\Upsilon}(\bar{z}_0-\bar{z})}\right), \tag{3.14}$$

where the following is taken into account

$$\int\limits_{1+e^{\bar{\Upsilon}\bar{z}_0}}^{1+e^{\bar{\Upsilon}(\bar{z}_0-\bar{z})}}\frac{\ln y}{1-y}dy=-\int\limits_{1+e^{\bar{\Upsilon}(\bar{z}_0-\bar{z})}}^{1+e^{\bar{\Upsilon}\bar{z}_0}}\frac{\ln y}{1-y}dy=-\int\limits_{1}^{1+e^{\bar{\Upsilon}\bar{z}_0}}\frac{\ln y}{1-y}dy+\int\limits_{1}^{1+e^{\bar{\Upsilon}(\bar{z}_0-\bar{z})}}\frac{\ln y}{1-y}dy=-\mathrm{Li}_2\left(-e^{\bar{\Upsilon}\bar{z}_0}\right)+\mathrm{Li}_2\left(-e^{\bar{\Upsilon}(\bar{z}_0-\bar{z})}\right),$$

$$\mathrm{Li}_2\left(1-z\right)\overset{\det}{=}\int\limits_1^z\frac{\ln s}{1-s}ds. \tag{3.15}$$

Function $\mathrm{Li}_2$ (3.15) is a special function, a *dilogarithm*. Substituting integral (3.14) into expression (3.13), we obtain a representation for a partial solution (3.12):

$$\begin{aligned}u_{p.i.}^{(1)}\left(z\right)=&-\frac{\bar{\alpha}_T}{2d_0}z^2-\frac{\bar{\alpha}_T L}{d_0\bar{\Upsilon}}\ln\frac{1+e^{-\Upsilon z_2}}{1+e^{\Upsilon z_1}}z+\frac{\bar{\alpha}_T L^2}{d_0\bar{\Upsilon}^2}\left[\mathrm{Li}_2\left(-e^{\Upsilon z_1}\right)+\mathrm{Li}_2\left(-e^{-\Upsilon z_2}\right)\right]-\\&-\frac{\bar{\alpha}_T L^2}{d_0\bar{\Upsilon}^2}\left[\mathrm{Li}_2\left(-e^{-\Upsilon(z_2-z)}\right)+\mathrm{Li}_2\left(-e^{\Upsilon(z_1-z)}\right)\right].\end{aligned} \tag{3.16}$$

Expression (3.16) consists of two parts: parabolic and dilogarithmic. Since solution (3.16) must satisfy equation (3.12), the linear part in expression (3.16) can be ignored, i.e.

$$u_{p.i.}^{(2)}\left(z\right)=-\frac{\bar{\alpha}_T}{2d_0}z^2-\frac{\bar{\alpha}_T L^2}{d_0\bar{\Upsilon}^2}\left[\mathrm{Li}_2\left(-e^{-\Upsilon(z_2-z)}\right)+\mathrm{Li}_2\left(-e^{\Upsilon(z_1-z)}\right)\right]. \tag{3.17}$$

Both functions (3.16) and (3.17) are partial solutions of equation (2.41), but they satisfy different initial-boundary conditions:

$$u_{p.i.}^{(1)}\left(0\right)=0,\quad u_{p.i.}^{(2)}\left(0\right)=-\frac{\bar{\alpha}_T L^2}{d_0\bar{\Upsilon}^2}\left[\mathrm{Li}_2\left(-e^{-\Upsilon z_2}\right)+\mathrm{Li}_2\left(-e^{\Upsilon z_1}\right)\right]. \tag{3.18}$$

The difference (3.18) is due to the neglect of linear terms in solution (3.17). On the one hand, if we assume that $\tau\left(0\right)\approx\tau''\left(1\right)\approx\tau'''\left(1\right)\approx 0$ (within the approximation (1.8)), then partial solution (3.16) will satisfy the boundary conditions on the function itself $u\left(z,t\right)$. Consequently, after a sufficient period of time, solution $u_{g.h}\left(z,t\right)$ containing damped oscillations will become negligibly small compared to particular solution $u_{p.i.}^{(1)}\left(z\right)$. Thus, the following approximation will be valid

$$\lim_{t\to+\infty}u\left(z,t\right)=u_{p.i.}^{(1)}\left(z\right). \tag{3.19}$$

Our solution $u(z,t)$ is known to be represented as the following sum (2.1) $u(z,t)=u_{g.h.}(z,t)+u_{p.i.}^{(1)}(z)$, where $u_{g.h.}(z,t)$ is the general solution for the homogeneous equation, and $u_{p.i.}^{(1)}(z)$ is the partial solution of the inhomogeneous equation. The $u_{g.h.}(z,t)$ solution is damping over time, which can be seen in (2.5), that is $\lim_{t\to+\infty} u_{g.h.}(z,t)=0$. Thus, the expression (3.19) is valid. From the physical point of view the oscillation of $u_{g.h.}(z,t)$, induced by the heat shock wave, will damp with time leaving just a static bend $u_{p.i.}^{(1)}(z)$, induced by the uneven heating of the rod walls.

On the other hand, for the time independent problem of thermal deformation of a fuel element, $\langle T\rangle(z,t)=\beta(t)\tau(z)=T_0=const$ solution of equation $u_{zzzz}=\overline{f}(z)$, (1.13) has the form:

$$u(z)=-\frac{\alpha_T \iota T_0}{2d_0}z^2. \qquad (3.20)$$

Under the condition $\iota\beta(t)=1^{\circ}K$ it follows that $\iota=1^{\circ}K/\beta=1^{\circ}K/T_0$ (since $\tau(z)=1$) and solution (3.20) goes into the following

$$u^{(3)}(z)=-\frac{\overline{\alpha}_T}{2d_0}z^2. \qquad (3.21)$$

Note that solutions (3.21) and (3.19), (3.16) do not coincide. This difference is due to the type of scaling function $\tau(z)$. In the case of solution (3.16), function $\tau(z)$ has dependence (1.8), while for solution (3.20) it is a constant. If we approximate function $\overline{\tau}(\overline{z})=\mathrm{rect}\left[\frac{2\overline{z}-\overline{z}_2-\overline{z}_1}{2(\overline{z}_2-\overline{z}_1)}\right]$, where rect is a rectangular function (see Fig. 3), then solution (3.21) can be rewritten as follows:

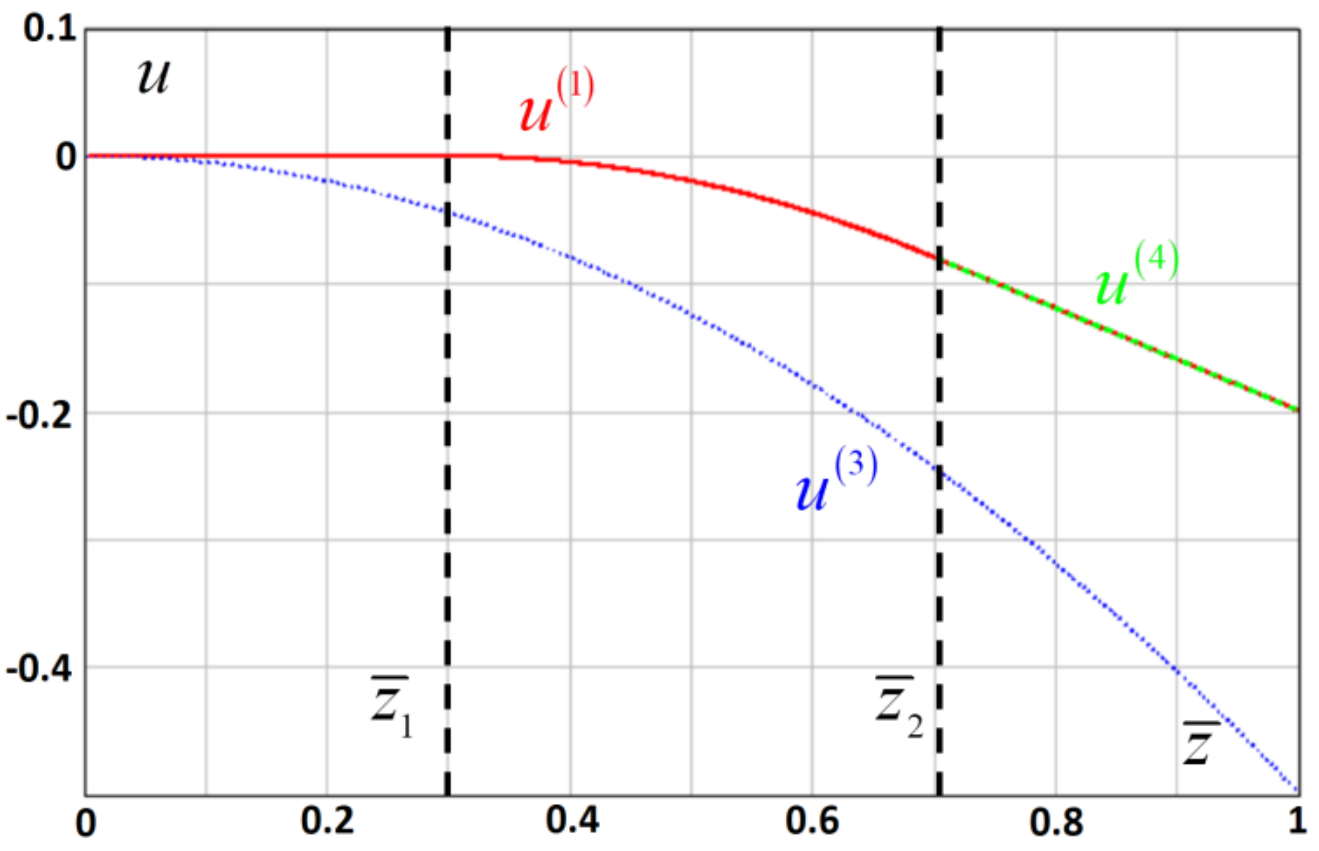

Fig. 9 Comparison of solutions with different approximations $\tau(\overline{z})$

$$u^{(4)}(\overline{z})=-\frac{\overline{\alpha}_T L^2}{2d_0}\begin{cases}(\overline{z}-\overline{z}_1)^2\,\mathrm{rect}\left[\frac{2\overline{z}-\overline{z}_2-\overline{z}_1}{2(\overline{z}_2-\overline{z}_1)}\right], & \overline{z}<\overline{z}_2,\\ (\overline{z}_2-\overline{z}_1)(2\overline{z}-1), & \overline{z}>\overline{z}_2,\end{cases} \qquad \mathrm{rect}(\overline{z})=\begin{cases}0, & |\overline{z}|>1/2,\\ 1/2, & |\overline{z}|=1/2,\\ 1, & |\overline{z}|<1/2,\end{cases} \qquad (3.22)$$

where it is assumed that $\overline{z}_1+\overline{z}_2=1$. Fig. 9 shows the graphs of functions (3.22), (3.21) and (3.16). It can be seen that solution (3.22) (green) practically coincides with solution (3.19), (3.16) (red), while solution (3.30) (blue) differs significantly. Rather complex functional dependence (3.16) is approximated with good accuracy by simplest function (3.22). As noted above, this result is predictable and is due to different approximations $\tau(z)$.

## §4 Simulation results

Let us consider examples of numerical simulation of the processes described in §1-3. The driving force on the right-hand side of equation (1.17) is determined by two functions $\beta(t)$ and $\tau(z)$, which are set before starting the calculations. The general view of function $\tau(z)$ is shown in Fig. 3, and its functional dependence can be approximated by expression (1.8) or, for example, by a rectangular function (see §3). From a practical point of view, an important case of dependence $\beta(t)$ is a rectangular function:

$$\beta(t)=\beta_0 \operatorname{rect}\left(\frac{2t-\Delta t}{2\Delta t}\right), \qquad \operatorname{rect}(t)=\theta\left(t+\frac{1}{2}\right)-\theta\left(t-\frac{1}{2}\right), \tag{4.1}$$

where $\Delta t$ is the duration of the thermal pulse of amplitude $\beta_0$; $\theta(t)$ is the Heaviside function. Functions (4.1) can be used to construct a periodic sequence of pulses with different relative duration. Further simulation in this section will be carried out with functions of type (4.1).

As shown in §2-3, the solution of problem (1.17)-(1.20) is reduced to the consideration of two main cases: $\lambda=0$ and $\lambda\neq 0$. Each of these cases contains various private options depending on the value of the parameter $p$. In fact, the values of parameter $p$ determine the form of the coordinate eigenfunctions in terms of which the solution of the original problem (1.17)-(1.20) will be expanded.

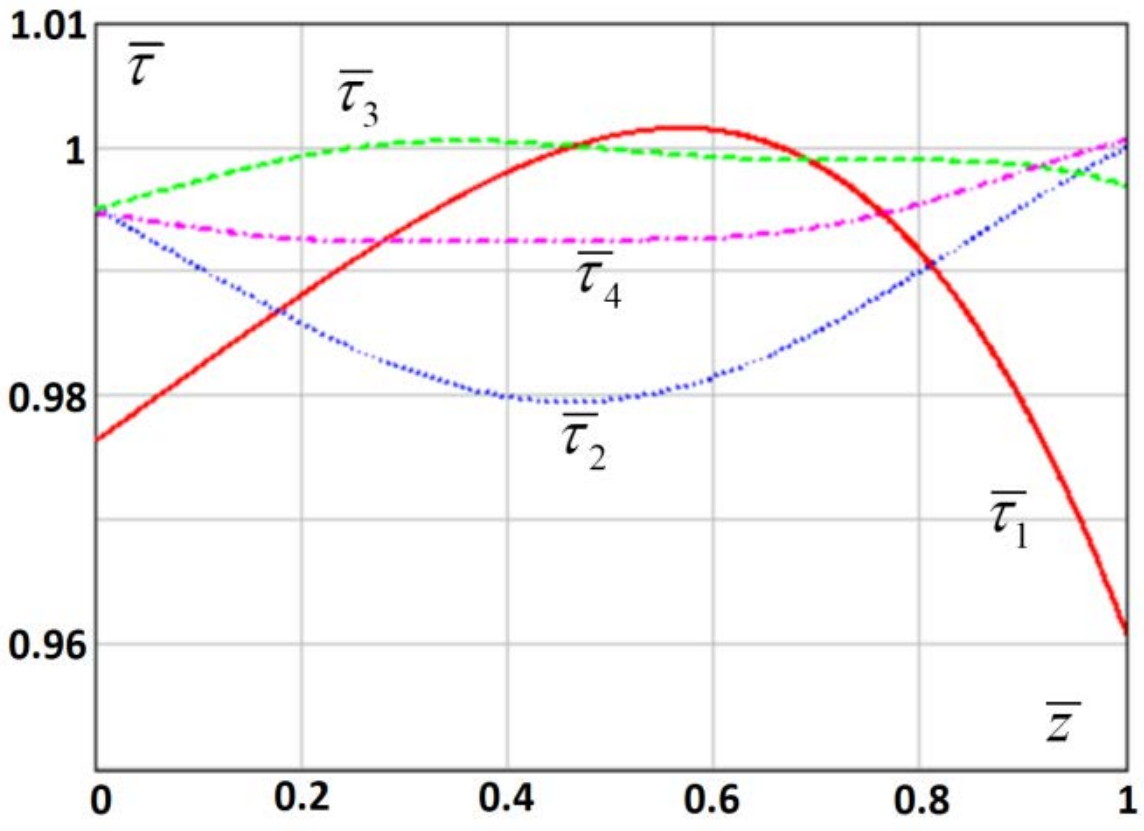

Fig. 10 Distribution of $\bar{\tau}(\bar{z},\ell)$

Let us start with the case $\lambda=0$ and the values of parameter $p=p_\ell$ for eigenfunctions $\bar{Z}_\ell^{(0)}(\bar{z})$. As shown in §2, distribution functions $\bar{\tau}(\bar{z},\ell)$ have a certain form (2.18). Fig. 10 shows four distributions of $\bar{\tau}(\bar{z},\ell)$ for $\ell=1...4$ with different sets of constants $\tau_j^{(\ell)}$, $j=0...2$. The values of constants $\tau_j^{(\ell)}$ in Fig. 10 are chosen in such a way that on average (with a deviation of no more than 0.05) function $\bar{\tau}(\bar{z},\ell)$ does not differ from unity. The functions in Fig. 10 correspond to a uniform average temperature distribution along the entire length of the rod. As an example for the case $\lambda=0$, we consider a simplified function (4.1), namely

$$\beta(t)=\beta_0\theta(\Delta t-t), \tag{4.2}$$

where $\theta$ is the Heaviside function. This simplification is made to remove the singularity in the initial conditions of the boundary value problem (1.17). When working with non-smooth functions, it is necessary to use the mathematical apparatus of the theory of generalized functions, which will be applied at $\lambda\neq 0$. As a result, according to (2.20), the solution of the boundary value problem (1.17) has the form:

$$u_\ell(z,t)=\bar{Z}_\ell^{(0)}(\bar{z})\,\chi(t), \tag{4.3}$$

$$\chi(t)=-\frac{\alpha_T \iota\tau_0}{d_0 L^2 a_1}\int_0^t e^{-2\gamma\bar{t}}d\bar{t}\int_0^{\bar{t}}\beta(s)e^{2\gamma s}ds+\frac{\alpha_T \iota\tau_0\kappa_0}{d_0 L^2 2\gamma}\left(e^{-2\gamma t}-1\right)+const. \tag{4.4}$$

The form of function $\beta(t)$ in integral (4.4) determines the evolution of the system. Expression (4.4) consists of two summand. The second summand determines the exponential «cooling» of the rod, which tends to zero at $t \to +\infty$. The first summand in expression (4.4) is related to the impact of external heat sources on the rod through function $\beta(t)$. If there are no heat sources ($\beta(t) \equiv 0$), then the «cooling» of the rod is observed (the second summand in (4.4)). Let us find the integral in expression (4.4) under condition (4.2), we obtain:

$$\chi(t) = -\frac{\alpha_T \iota \tau_0}{2d_0 L^2 \gamma a_2}\left[\beta_0 \Omega(t) + a_2 \kappa_0 \left(1 - e^{-2\gamma t}\right)\right] + 1, \tag{4.5}$$

$$\Omega(t) = \begin{cases} 2\gamma\Delta t + \left(e^{-2\gamma\Delta t} - 1\right)\left(1 + e^{-2\gamma\Delta t} - e^{-2\gamma t}\right), & t \ge \Delta t, \\ 2\gamma t + e^{-2\gamma t} - 1, & 0 \le t < \Delta t, \end{cases}$$

here unity was taken as the free integration constant. It follows from expression (4.5) that at the initial moment of time (before the start of the pulse) $\chi(0) = 1$, that is, the initial strain of the rod (4.3) coincides with eigenfunction $u_\ell(z,0) = \overline{Z}_\ell^{(0)}(\overline{z})$.

The velocity distribution at the initial moment of time is also proportional to the eigenfunction $\frac{\partial}{\partial t} u_\ell(z,0) = \chi'(0)\overline{Z}_\ell^{(0)}(\overline{z})$. During the pulse ($0 < t < \Delta t$) according to (4.5) there is a continuous increase in thermal strain. After the end of the pulse ($t > \Delta t$), there is a gradual deceleration of the strain rate and an *infinitely long* transition to a static strained state. Let us illustrate the described physical process.

Let us first consider a continuous thermal pulse $\Delta t \gg 1$. Fig. 11 shows the graphs of the evolution (at times $0 = t_0 < t_1 < t_2 < t_3 < t_4$) of the shape of the rod geometry for different eigenvalues $p_\ell$. The calculation data are given in relative dimensionless units. The arrows in Figs. 11, 12 show the direction of heat outflow from the fuel element. In all shown in Fig. 11 cases, there is no oscillation process, but the state with an eigenvalue $p_1$ is fundamentally different from the states with $p_2$, $p_3$, $p_4$. The difference consists in the presence of nodal points in states $p_2$, $p_3$, $p_4$, and, accordingly, their absence in state $p_1$. As shown earlier, such behavior is caused by the presence of a bifurcation point in the transition from the eigenvalue $p_1$ to $p_2$ and so on.

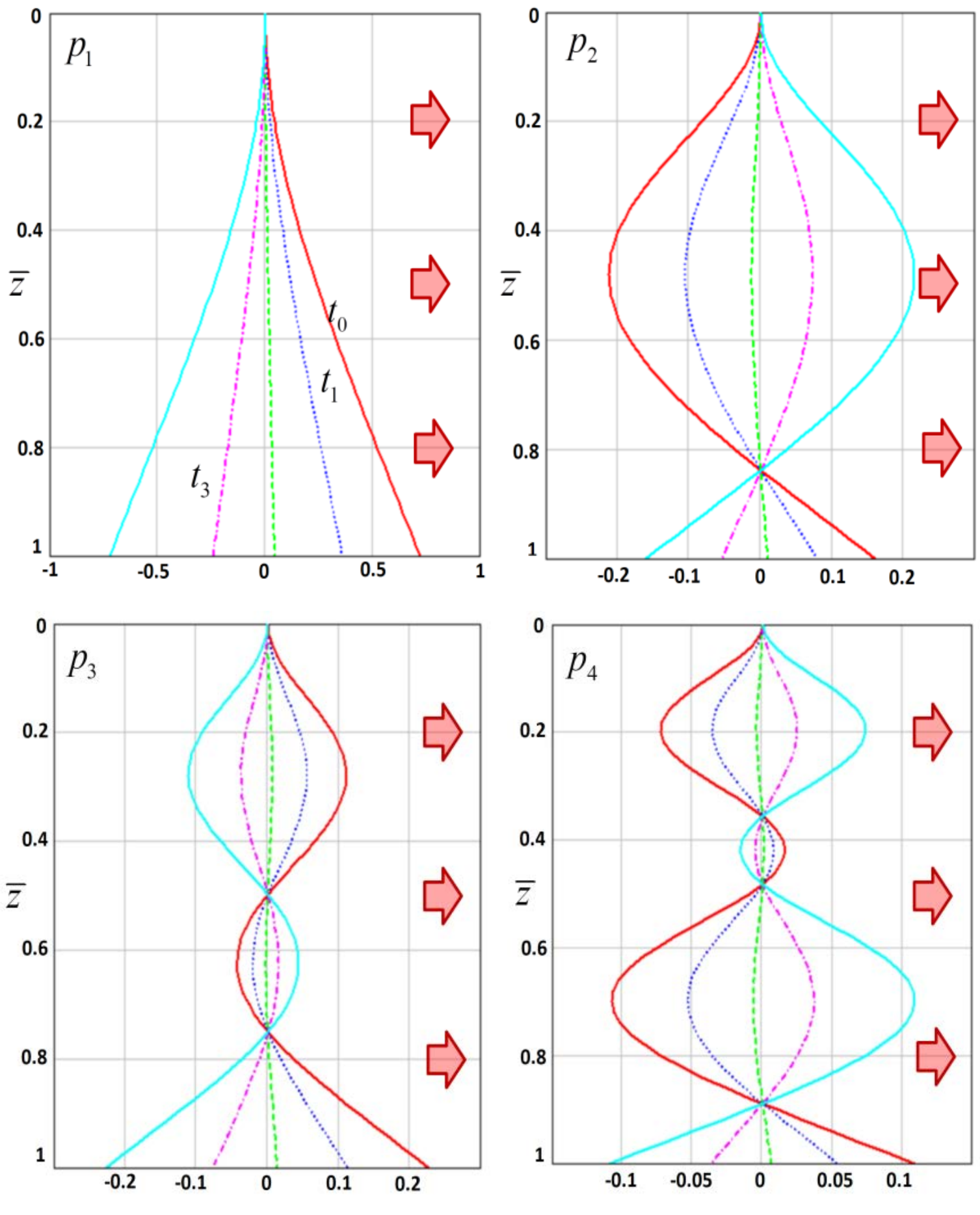

Fig. 11 Dynamics of the rod at $\lambda = 0$, $p = p_\ell$ at $\Delta t \gg 1$

For state $p_1$ at the initial moment of time, the rod is bent towards the outflow of heat (on the right). As a result of surface $S_+$ heating, it first straightens, and then bends in the opposite direction (to the left). With a constant continuous heating process, the strain in the form of a rod bend to the left will only intensify. A similar process occurs for states $p_2$, $p_3$, $p_4$, but there are nodal points that, despite continuous heating, remain motionless.

In the case of a short-term thermal pulse $\Delta t \sim 1$, the evolution of the system for various eigenvalues $p_\ell$ is shown in Fig. 12. As in the previous case, there is a bifurcation effect here. Pulse duration $\Delta t$ is chosen so that the asymptotic behavior of the final state of the rod corresponds to the undeformed state and $\Delta t < t_1$. As in the previous case ($\Delta t \gg 1$), an active deformation process is observed at the beginning of the dynamics ($0 < t < \Delta t$).

After the end of pulse $t > \Delta t$, the process of thermal strain damps exponentially, and as a result, the rod tends to the statically deformed state for an infinitely long time, since $\lim\limits_{t\to+\infty} \chi'(t) = 0$. Indeed, at long times ($t \gg 1$):

$$\chi'(t) = -\frac{\alpha_T l \tau_0}{d_0 L^2 a_2}\left[\beta_0\left(e^{-2\gamma\Delta t} - 1\right) + a_2\kappa_0\right]e^{-2\gamma t}, \quad t > \Delta t, \tag{4.6}$$

$$\chi''(t) = 2\gamma\frac{\alpha_T l \tau_0}{d_0 L^2 a_2}\left[\beta_0\left(e^{-2\gamma\Delta t} - 1\right) + a_2\kappa_0\right]e^{-2\gamma t},$$

where $\chi''(t) \ll 1$ is proportional to the average force acting on the rod and is related to the stiffness $EJ_x$.

An infinitely long movement in one direction (in this case from right to left) is associated with large eigenvalues $p_\ell = \dfrac{L^2 M_0 g}{EJ_x}$, the reason for which, from a physical point of view, is large mass $M_0$ and low stiffness $EJ_x$. As a result, the frictional force with damping coefficient $\gamma$ leads to «oscillations» with an infinite period.

Let us consider the case when $\lambda \neq 0$, and the right-hand side of equation (1.17) is determined by function (4.2) and (1.8). In this case, for $t < \Delta t$, equation (1.17) is inhomogeneous, and for $t \geq \Delta t$, it will be homogeneous. The construction of partial solution $u_{p.i.}$ of an inhomogeneous equation (for the right-hand side case under consideration) is described in detail in §3. It is shown that a partial solution can be chosen to be time independent (3.16), while it admits the simplest approximation (3.22) (see Fig. 10). Approximation (3.22) is especially important in numerical simulation, as it significantly reduces the calculation time. Thus, in the case of function (4.2), the time dependent partial solution of equation (1.14) can be represented as:

$$u_{p.i.}(\overline{z},t)=u_{p.i.}^{(1)}(\overline{z})\theta(\Delta t-t) \quad \text{or} \quad u_{p.i.}(\overline{z},t)=u_{p.i.}^{(4)}(\overline{z})\theta(\Delta t-t), \tag{4.7}$$

where the second variant in expression (4.7) is applicable for numerical simulation. General solution (2.1) according to (2.68) has the form

$$u(z,t)=\sum_{n=1}^{+\infty}e^{-\gamma t}\left[A_n\cos(\beta_n t)+B_n\sin(\beta_n t)\right]\Phi_n(\overline{z})+u_{p.i.}(\overline{z},t), \tag{4.8}$$

where the form of eigenfunctions $\Phi_n(\overline{z})$ is determined by the value of parameter $p$ (see §2-3). Coefficients $A_n$ and $B_n$ in expansion (4.8) are determined in accordance with the initial conditions (2.70):

$$\begin{gathered}\sum_{n=1}^{+\infty}u_n^{(0)}\Phi_n(\overline{z})-u_{p.i.}^{(1)}(\overline{z})=\sum_{n=1}^{+\infty}A_n\Phi_n(\overline{z}),\\ \sum_{n=1}^{+\infty}u_n^{(1)}\Phi_n(\overline{z})=\sum_{n=1}^{+\infty}(\beta_n B_n-\gamma A_n)\Phi_n(\overline{z}).\end{gathered} \tag{4.9}$$

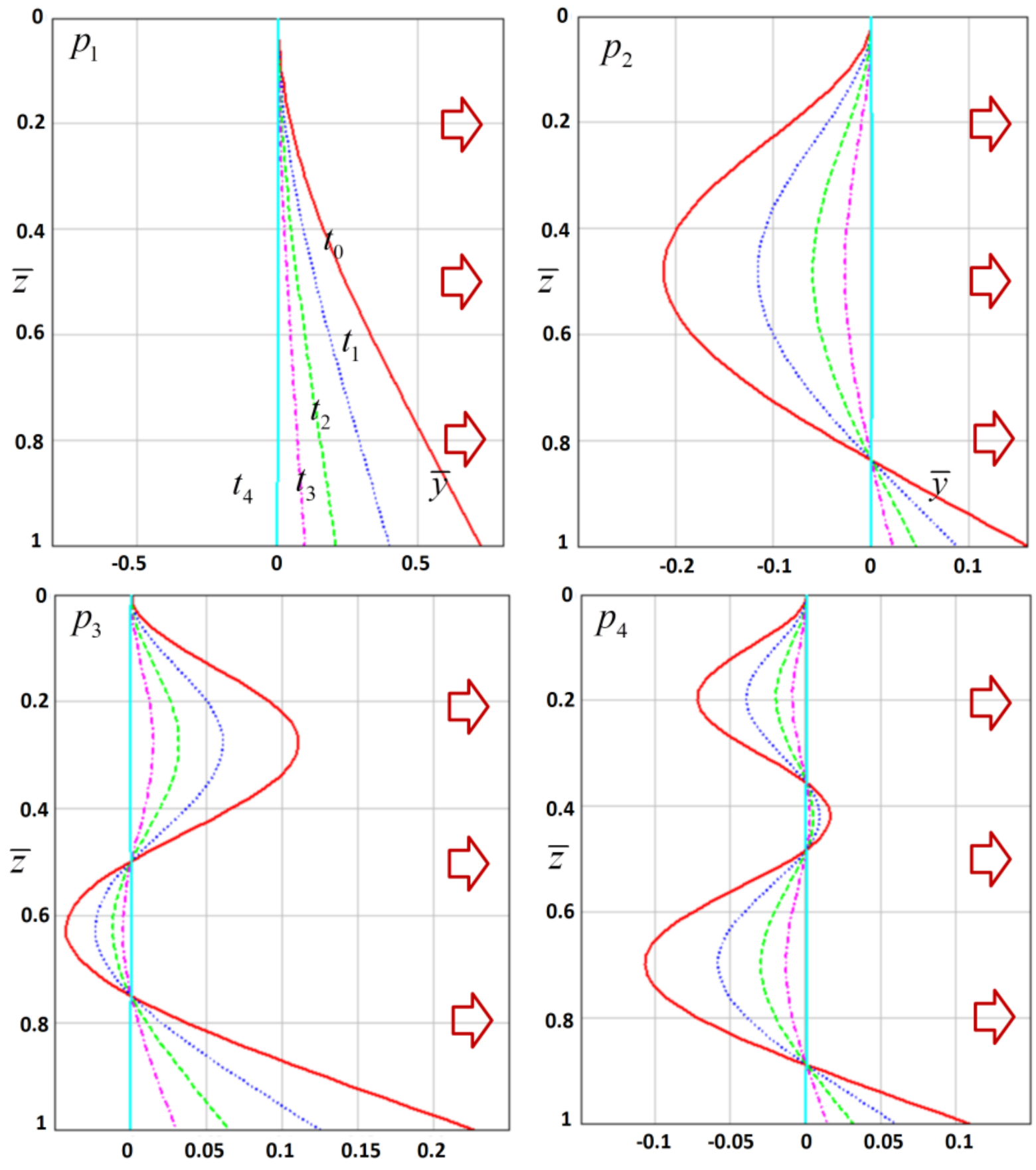

Fig. 12 Dynamics of the rod at $\lambda = 0$, $p = p_\ell$ at $\Delta t \sim 1$

Taking into account the orthogonality of functions $\Phi_n(\overline{z})$, expressions (4.9) take the form:

$$A_k=u_k^{(0)}-\frac{1}{\|\Phi_k\|^2}\int_0^1 u_{p.i.}^{(1)}(\overline{z})\Phi_n(\overline{z})d\overline{z}, \qquad B_k=\frac{u_k^{(1)}+\gamma A_k}{\beta_k}. \tag{4.10}$$

To simplify the calculation of the integral in expression (4.10), one can use the approximation $u_{p.i.}^{(4)}(\overline{z})$. Substitution of values (4.10) into expression (4.8) gives the final form of the solution.

As an example, consider the simplest form of eigenfunctions when $p=0$ (see §3), that is, $\Phi_k=W_k$. The form of functions $W_k$ allows one to find the integral (4.10) with function $u_{p.i.}^{(4)}(\overline{z})$ explicitly:

$$\begin{gathered}\frac{1}{\|W_n\|^2}\int_0^1 u_{p.i.}^{(4)}(\overline{z})W_n(\overline{z})d\overline{z}=-\frac{\overline{\alpha}_T L^2}{2d_0}\left[I_n^{(1)}(\overline{z}_1,\overline{z}_2)+(\overline{z}_2-\overline{z}_1)I_n^{(2)}(\overline{z}_2)\right],\\ I_n^{(1)}(\overline{z}_1,\overline{z}_2)=\int_{\overline{z}_1}^{\overline{z}_2}(\overline{z}-\overline{z}_1)^2W_n(\overline{z})d\overline{z},\ I_n^{(2)}(\overline{z}_2)=\int_{\overline{z}_2}^{1}(2\overline{z}-1)W_n(\overline{z})d\overline{z},\end{gathered} \tag{4.11}$$

where integrals $I_n^{(1)}$ and $I_n^{(2)}$ are expressed in terms of the Krylov functions (3.2)

$$I_n^{(2)}(\overline{z}_2)=\frac{2}{\alpha_n}(2\overline{z}_2-1)\mathrm{K}_4(\alpha_n)\left[\frac{\mathrm{K}_1(\alpha_n\overline{z}_2)}{\mathrm{K}_1(\alpha_n)}-\frac{\mathrm{K}_4(\alpha_n\overline{z}_2)}{\mathrm{K}_4(\alpha_n)}\right]+$$
$$+\frac{4}{\alpha_n^2}\mathrm{K}_1(\alpha_n)\left[\frac{\mathrm{K}_1(\alpha_n\overline{z}_2)}{\mathrm{K}_1(\alpha_n)}-\frac{\mathrm{K}_2(\alpha_n\overline{z}_2)}{\mathrm{K}_2(\alpha_n)}\right], \tag{4.12}$$
$$I_n^{(1)}(\overline{z}_1,\overline{z}_2)=\frac{4}{\alpha_n^3}\mathrm{K}_1(\alpha_n)\left[\frac{\mathrm{K}_2(\alpha_n\overline{z}_2)-\mathrm{K}_2(\alpha_n\overline{z}_1)}{\mathrm{K}_1(\alpha_n)}-\frac{\mathrm{K}_3(\alpha_n\overline{z}_2)-\mathrm{K}_3(\alpha_n\overline{z}_1)}{\mathrm{K}_2(\alpha_n)}\right]+$$
$$+\frac{4(\overline{z}_1-\overline{z}_2)}{\alpha_n^2}\mathrm{K}_1(\alpha_n)\left[\frac{\mathrm{K}_1(\alpha_n\overline{z}_2)}{\mathrm{K}_1(\alpha_n)}-\frac{\mathrm{K}_2(\alpha_n\overline{z}_2)}{\mathrm{K}_2(\alpha_n)}\right]+\frac{2(\overline{z}_2-\overline{z}_1)^2}{\alpha_n}\mathrm{K}_4(\alpha_n)\left[\frac{\mathrm{K}_4(\alpha_n\overline{z}_2)}{\mathrm{K}_4(\alpha_n)}-\frac{\mathrm{K}_1(\alpha_n\overline{z}_2)}{\mathrm{K}_1(\alpha_n)}\right].$$

Using explicit expressions (4.10)-(4.12) it is possible to efficiently calculate expansion coefficients $A_n$ and $B_n$ with high accuracy. Thus, solution (4.8) has been completely found explicitly. Figs. 13, 14 show the results of numerical simulation of solution (4.8). The following parameters were taken as initial parameters (see Fig. 1):

- *geometric parameters:*

$$L=1m,\quad d_0=0.017m,\quad \Delta d_0=4.5\cdot 10^{-4}m, \tag{4.13}$$
$$J_x=\frac{\pi}{4}\left(R_2^4-R_1^4\right),\ R_2=\frac{d_0}{2},\ R_1=R_2-\Delta d_0,$$

- *material and medium properties:*

$$\rho=7950\frac{kg}{m^3},\quad E=200\,GPa,\quad \alpha_T=17.3\cdot 10^{-6}\frac{1}{^\circ C},\quad \eta=11.24\frac{s\cdot N}{m^3},\quad \iota=0.05, \tag{4.14}$$

- *thermal pulse (4.2):*

$$\Delta t=1s,\qquad \iota\beta_0=1^\circ K. \tag{4.15}$$

Note that the value of parameter $p\approx 0.011$ corresponds to the data (4.13), (4.14), that is eigenfunctions $\Phi_n(\overline{z},p)$ almost completely coincide with functions $W_n(\overline{z})$.

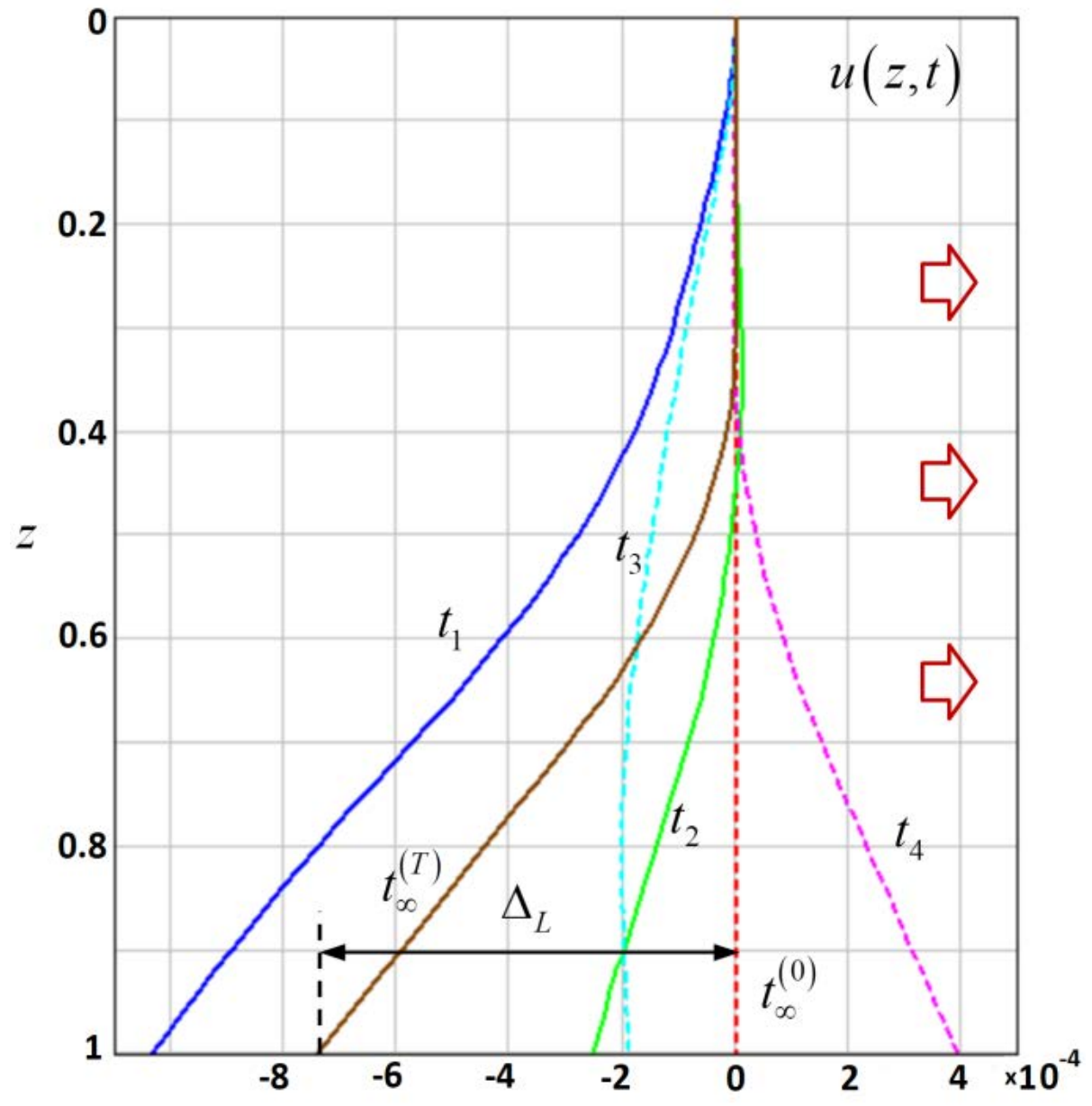

Fig. 13 Effect of pulse on the dynamics of the rod

The system has zero initial conditions, that is $u_k^{(0)}=u_k^{(1)}=0$ (4.10). During a period of time $t<\Delta t$ an external force in the form of a thermal pulse acts on the rod (red arrows on the right in Fig. 13 indicate the direction of heat outflow). The thermal pulse shifts the «balance» point, around which the rod tip oscillates, to the left by a distance $\Delta_L=\left|u_{p.i.}^{(4)}(L)\right|$. Fig. 13 shows two main centers of «equilibrium» ($t_\infty^{(T)}$ and $t_\infty^{(0)}$), corresponding to the asymptotic of the solution at $t\to+\infty$ under the constant action of a thermal pulse ($t_\infty^{(T)}$) and in its absence ($t_\infty^{(0)}$). The shape of the rod for $t_\infty^{(T)}$ corresponds to time independent partial solution (3.31), and the shape of the rod for $t_{\infty}^{(0)}$ corresponds to the state of rest without the action of external forces. The rod shapes at time moments $t_1, t_2$ (solid line in Fig. 13) characterize the oscillation amplitude at a certain time moment $0 < t < \Delta t$, and the rod shapes at moments $t_3, t_4$ (dotted line in Fig. 13) characterize time moment $t > \Delta t$. The presence of a friction force leads to exponential damping of oscillations with coefficient $\gamma$ over the entire time interval (see Fig. 14).

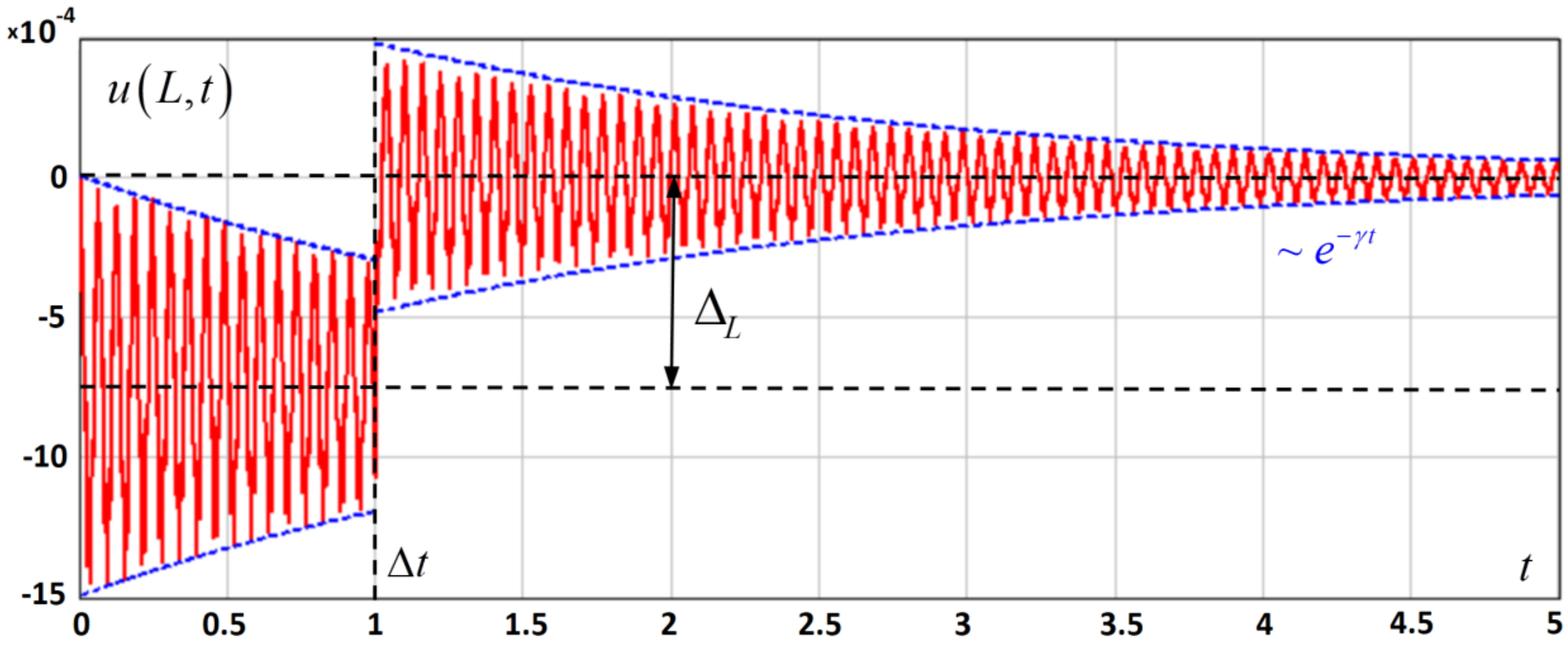

Fig. 14 Evolution of the rod end motion during the pulse time and after it

Fig. 14 shows the trajectory of the rod tip over time. As mentioned earlier, in period of time $0 < t < \Delta t$, the center of equilibrium of the oscillation system (the rod tip) is shifted by distance $\Delta_L$ due to the continuous action of the heat source. At the moment of time $t = \Delta t$, the action of the source stops, and the center of equilibrium moves to the origin of coordinates ( $x = y = 0$, see Figs. 14, 1). Further motion at $t > \Delta t$ corresponds to free damped oscillations (see Fig. 14).

The considered type of motion with zero initial conditions was caused by an initially acting external force. The force was present not only on time interval $0 < t < \Delta t$, but also at $t \le 0$. In real physical systems, a different situation may arise. At initial moment of time $t = 0-$, the system has zero initial conditions, and at moment of time $t = 0+$, the conditions «instantly» change:

$$u\big|_{t=0+} = u_0(z), \qquad \left.\frac{\partial u}{\partial t}\right|_{t=0+} = u_1(z), \tag{4.16}$$

but at the same time

$$u\big|_{t=0-} = 0, \qquad \left.\frac{\partial u}{\partial t}\right|_{t=0-} = 0. \tag{4.17}$$

Conditions (4.17) mean that «before» the initial moment of time ($t = 0-$) the rod was not deformed and was at rest, and «after», at initial moment $t = 0+$, the rod was instantly heated, which led to non-zero conditions (4.16). As noted above, the abrupt change in the initial conditions is caused by an external force defined through function (4.1). In this situation, the construction of a solution to equation (2.41) requires the use of the mathematical apparatus of the theory of generalized functions.

Let us construct fundamental solution $\mathcal{E}_n(t) = \mathcal{W}_n(t)\theta(t)$ of equation (2.41). For convenience, we define operator $\bar{\mathfrak{L}}_n$: $\bar{\rho} a_1 \bar{\mathfrak{L}}_n = \mathfrak{L}_n$, $a_1 \bar{\mathrm{f}}_n(t) = \bar{f}_n(t)$ (2.50):

$$\begin{cases} \bar{\mathfrak{L}}_n \bar{w}_n(t) = \bar{\mathrm{f}}_n(t), \\ \bar{w}_n(0) = \bar{w}_n^{(0)}, \ \bar{w}_n'(0) = \bar{w}_n^{(1)}, \end{cases} \tag{4.18}$$

$$\bar{\mathfrak{L}}_n \mathcal{E}_n(t) = \delta(t), \qquad \begin{cases} \bar{\mathfrak{L}}_n \mathcal{W}_n = 0, \\ \mathcal{W}_n(0) = 0, \ \mathcal{W}_n'(0) = 1, \end{cases} \tag{4.19}$$

where coefficients $\bar{w}_n^{(0)}$, $\bar{w}_n^{(1)}$ in problem (4.18) are determined through the initial conditions (4.16), that is, $\bar{w}_n^{(0)} = u_n^{(0)}$, $\bar{w}_n^{(1)} = u_n^{(1)}$. The solution of the Cauchy problem (4.19) has the form:

$$\mathcal{W}_n(t) = \frac{1}{\beta_n} e^{-\gamma t} \sin(\beta_n t), \ \ \mathcal{E}_n(t) = \frac{1}{\beta_n} \theta(t) e^{-\gamma t} \sin(\beta_n t). \tag{4.20}$$

Let us extend the Cauchy problem to domain $t < 0$. For the right-hand side of equation (4.18) and function $\bar{w}_n$, we obtain:

$$\tilde{f}_n(t) = \bar{f}_n(t)\theta(t), \tag{4.21}$$

$$\tilde{w}_n'(t) = \{\bar{w}_n'(t)\} + \bar{w}_n^{(0)}\delta(t), \ \ \tilde{w}_n''(t) = \{\bar{w}_n''(t)\} + \bar{w}_n^{(0)}\delta'(t) + \bar{w}_n^{(1)}\delta(t),$$

where $\{\bullet\}$ is the notation for the classical derivative. Taking into account expressions (4.21), equation (4.18) and its solution takes the form:

$$\bar{\mathfrak{L}}_n \tilde{w}_n(t) = \frac{1}{a_1}\tilde{f}_n(t) + \bar{w}_n^{(0)}\delta'(t) + \left[\bar{w}_n^{(1)} + \frac{a_2}{a_1}\bar{w}_n^{(0)}\right]\delta(t) \stackrel{\det}{=} F_n(t),$$

$$\tilde{w}_n(t) = (\mathcal{E}_n * F_n)(t). \tag{4.22}$$

Thus, it is necessary to calculate the convolution (4.22):

$$\tilde{w}_n(t) = (\mathcal{E}_n * F_n)(t) = \frac{1}{a_1}\int_0^t \bar{f}_n(\bar{t})\mathcal{W}_n(t-\bar{t})\,d\bar{t} + \bar{w}_n^{(0)}\mathcal{E}_n'(t) + \left(\bar{w}_n^{(1)} + \frac{a_2}{a_1}\bar{w}_n^{(0)}\right)\mathcal{E}_n(t). \tag{4.23}$$

As a result, at $t > 0$ (4.20) with taken into account, solution (4.23) takes the form:

$$a_1\beta_n\bar{w}_n(t) = \int_0^t \bar{f}_n(\bar{t})e^{-\gamma(t-\bar{t})}\sin\left[\beta_n(t-\bar{t})\right]d\bar{t} + a_1\bar{w}_n^{(0)}e^{-\gamma t}\left[\beta_n\cos(\beta_n t) - \gamma\sin(\beta_n t)\right] +$$

$$+\left(a_1\bar{w}_n^{(1)} + a_2\bar{w}_n^{(0)}\right)e^{-\gamma t}\sin(\beta_n t),$$

or

$$a_1\beta_n\bar{w}_n(t) = \bar{I}_n(t) + e^{-\gamma t}\left\{a_1\bar{w}_n^{(0)}\beta_n\cos(\beta_n t) + \left[a_1\left(\bar{w}_n^{(1)} - \bar{w}_n^{(0)}\gamma\right) + a_2\bar{w}_n^{(0)}\right]\sin(\beta_n t)\right\}, \tag{4.24}$$

$$\bar{I}_n(t) = \int_0^t \bar{f}_n(\bar{t})e^{-\gamma(t-\bar{t})}\sin\left[\beta_n(t-\bar{t})\right]d\bar{t}.$$

Since conditions (4.15) are considered, according to expression (3.8), function $\overline{f}_n$ is a constant value, therefore, integral (4.24) can be calculated explicitly:

$$\overline{I}_n(t) = -\frac{\overline{\alpha}_T \overline{\tau}''_n}{d_0 L^2} \int_0^t e^{-\gamma t'} \sin(\beta_n t')\,dt' = -\frac{\overline{\alpha}_T \overline{\tau}''_k}{d_0 L^2}\left[D_n(t) - D_n(0)\right],$$

$$D_n(t) = \int e^{-\gamma t} \sin(\beta_n t)\,dt = -e^{-\gamma t}\frac{\beta_n \cos(\beta_n t) + \gamma \sin(\beta_n t)}{\beta_n^2 + \gamma^2},$$

$$\overline{I}_n(t) = -\frac{\overline{\alpha}_T \overline{\tau}''_n}{d_0 L^2 (\beta_n^2 + \gamma^2)}\left\{\beta_n - e^{-\gamma t}\left[\beta_n \cos(\beta_n t) + \gamma \sin(\beta_n t)\right]\right\}. \quad (4.25)$$

Substituting (4.25) into (4.24) gives an explicit expression for coefficients $\overline{w}_n(t)$:

$$\overline{w}_n(t) = -\frac{\overline{\alpha}_T \overline{\tau}''_n}{a_1 d_0 L^2 (\beta_n^2 + \gamma^2)} + \quad (4.26)$$

$$+e^{-\gamma t}\left\{\left[\frac{\overline{\alpha}_T \overline{\tau}''_n}{a_1 d_0 L^2 (\beta_n^2 + \gamma^2)} + \overline{w}_n^{(0)}\right]\cos(\beta_n t) + \frac{1}{\beta_n}\left[\frac{\gamma \overline{\alpha}_T \overline{\tau}''_n}{a_1 d_0 L^2 (\beta_n^2 + \gamma^2)} + \overline{w}_n^{(1)} + \overline{w}_n^{(0)}\gamma\right]\sin(\beta_n t)\right\}.$$

Note that obtained solution (4.26) satisfies the initial conditions of the Cauchy problem (4.18) and (4.16). Coefficients $\overline{\tau}''_n$ are found by formulas (2.43). For numerical calculation, it is convenient to use the representation:

$$\overline{\tau}''_n = \int_0^1 \overline{\tau}(\overline{z}) W''_n(\overline{z})\,d\overline{z}, \quad (4.27)$$

$$W''_n(\overline{z}) = 2\alpha_n^2 \mathrm{K}_1(\alpha_n)\left[\frac{\mathrm{K}_1(\alpha_n \overline{z})}{\mathrm{K}_1(\alpha_n)} - \frac{\mathrm{K}_2(\alpha_n \overline{z})}{\mathrm{K}_2(\alpha_n)}\right],$$

where (3.2)-(3.5) are taken into account. Thus, the solution (2.39) has been constructed.

Let us consider three variants of the function $u_0(z)$ and $u_1(z)$ in the initial conditions (4.16):

$$u_0(z) = 0,\ u_1(z) = \varsigma \frac{u_{p.i.}^{(4)}(\overline{z})}{\delta t},\quad \varsigma = 0, 4, \quad (4.28)$$

where $\delta t$ is the «reaction» time of the rod material to a thermal agitation. We will focus on the dynamics of the rod tip, that is, we will build dependencies as in Fig. 14. For each value of coefficient $\varsigma$ (4.28) we calculate the solution (2.39).

Let us start with value $\varsigma = 0$, which corresponds to the initially resting rod. In contrast to the previous case, here the external force (thermal pulse) arises instantaneously (4.21) at $t > 0$ and continues throughout the observation period. Fig. 15 shows that in addition to damped oscillations, there are significant beats in time interval $0 < t < t_b = 1.5\,s$. The beat amplitude decreases with damping coefficient $\gamma$, while the beat frequency increases linearly with time $\varpi_0(t) = \omega_0 t + \varphi_0$.

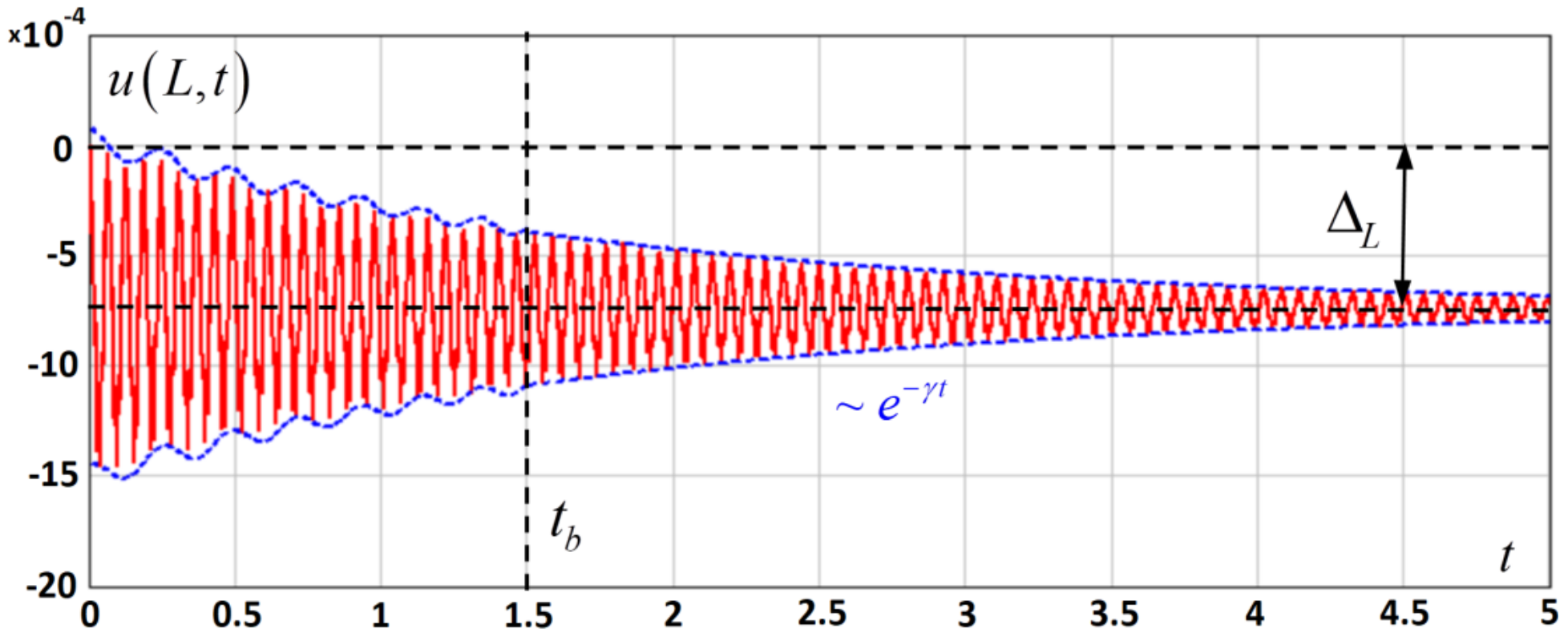

Fig. 15 Trajectory of the rod tip at instant pulse ( $\varsigma = 0$ )

From a comparison of Fig. 15 and Fig. 14 it follows that the physical nature of beats is the instantaneous action of force (4.21). Note that in Fig. 14, the beat pattern is partially observed after the termination of the force ($t > \Delta t$). At $t > t_b$, the beat amplitude drops significantly and ordinary damping oscillations set in around the «center of equilibrium» corresponding to the static thermal bending $y_0 = u_{p.i.}^{(4)}(L)$.

When $\varsigma = 4$ the initial velocity is distributed along the entire length of the rod according to (4.28). The choice of the velocity dependence (4.28) is due to the following considerations. If damping oscillations are not taken into account, then after heating the rod should take the form described by function $u_{p.i.}^{(4)}(z)$ in some characteristic time $\delta t$. Since there is no initial strain, the transition rate of each point of the rod is determined by value $u_{p.i.}^{(4)}(z)/\delta t$. The numerical value of $\delta t$ is a separate issue that requires additional study. From a physical point of view, it is interesting to study the influence of the initial velocity of the rod caused by thermal shock on the dynamics of the system. Such a choice of initial conditions (4.28) or (4.16)-(4.17) is suitable for the case when the thermal pulse hits the rod instantaneously, which is already oscillating in accordance with (4.16)-(4.17).

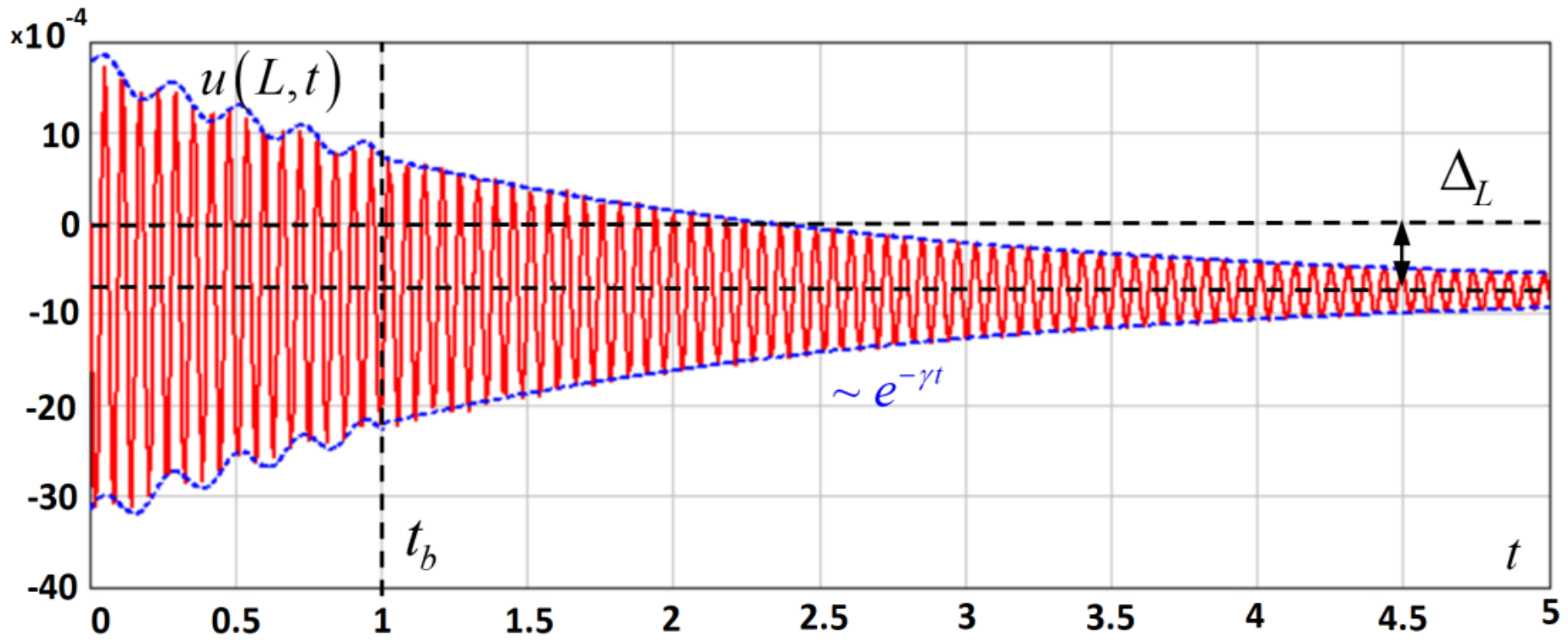

Fig. 16 Trajectory of the rod tip at instant pulse ( $\varsigma = 4$ )

Fig. 16 shows the dynamics of the rod tip at $\varsigma = 4$. Value $\Delta_L$ in all Figs. 14-16 is the same. Externally, the dynamics of the system behavior is similar to Fig. 15, but there are qualitative differences. First, the amplitude of oscillations has increased (see Fig. 16). This is the expected result, since the initial speed of the rod appeared. Secondly, the beat frequency has increased with almost the same amplitude. The approximate change in the beat frequency for the case $\varsigma = 4$ was about 5% ($\varpi_4(t) = 1.05\varpi_0(t)$). Third, there was a phase shift of the beats (see Figs. 15, 16). Fourthly, at $\varsigma = 4$, the beat amplitude drops significantly after $t > t_b = 1s$, that is, earlier than at $\varsigma = 0$ (see Figs. 15).

Let us consider in more detail the situation when the non-uniform heating of the rod occurs according to the periodic law. In this case, by the time of a new thermal pulse, the rod will have nonzero initial conditions in the Cauchy problem (4.16). From the examples discussed above, shown in Figs. 15, 16 ($\varsigma = 0, 4$), it can be seen that the presence of non-zero initial conditions leads to an increase in the oscillation amplitude. Consequently, under the action of a periodic force, a significant increase in the amplitude of the oscillations of the rod can occur.

We obtain an expression for solution (4.26) for the periodic law of the form:

$$\alpha_T \iota \beta(t) = \bar{\alpha}_T \Theta(t,\nu), \tag{4.29}$$

$$\Theta(t,\nu) = \sum_{k=0}^{N} \left[ \theta\big(k(\nu+1)\Delta t - t + \Delta t\big) + \theta\big(t - k(\nu+1)\Delta t\big) \right] - N - 1,$$

or

$$\Theta(t,\nu) = \begin{cases} 1, & k(\nu+1)\Delta t < t < k(\nu+1)\Delta t + \Delta t, \\ 0, & k(\nu+1)\Delta t + \Delta t < t < (k+1)(\nu+1)\Delta t + \Delta t, \end{cases}$$

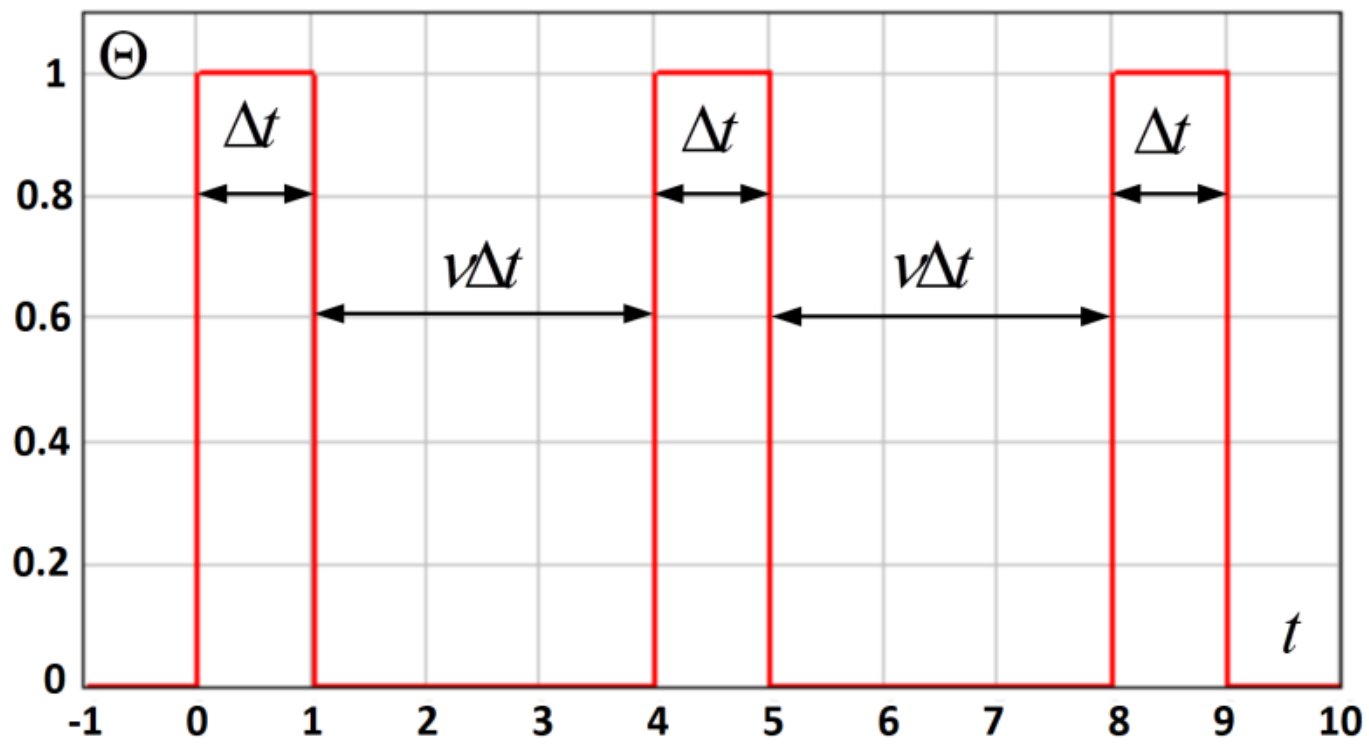

Fig. 17 Pulse function $\Theta(\nu = 3, t)$

where $k = 0,1,\ldots$; $N$ is the number of the pulses. Fig. 17 shows the graph of function $\Theta(t,\nu)$, which consists of periodic rectangular pulses of length $\Delta t$, spaced apart from each other by $\nu\Delta t$. Thus, the period is $(\nu+1)\Delta t$, and the pulses' duty cycle is $S = \nu + 1$.

In accordance with expressions (3.17) and (4.29), we obtain:

$$\bar{f}_k(t) = -\frac{\alpha_T \iota \bar{\tau}''_k}{d_0 L^2} \beta(t) = -\frac{\bar{\alpha}_T \bar{\tau}''_k}{d_0 L^2} \Theta(t,\nu). \tag{4.30}$$

We substitute the expression (4.30) into the integral (4.24) $\bar{I}_n(t)$, we obtain:

$$\bar{I}_n(t) = -\frac{\bar{\alpha}_T \bar{\tau}''_n}{d_0 L^2} \left\{ \sum_{k=0}^{N} \left[ I^{(1)}_{n,k}(t,\nu) + I^{(2)}_{n,k}(t,\nu) \right] - (N+1) I^{(3)}_n(t,\nu) \right\}, \tag{4.31}$$

$$I^{(1)}_{n,k}(t,\nu) = \begin{cases} D_n[t] - D_n\left[t - \Delta t - k\Delta t(\nu+1)\right], & t \ge \Delta t\left[1 + k(\nu+1)\right], \\ D_n[t] - D_n[0], & t < \Delta t\left[1 + k(\nu+1)\right], \end{cases}$$

$$I^{(2)}_{n,k}(t,\nu) = \begin{cases} D_n\left[t - k(\nu+1)\Delta t\right] - D_n[0], & t \ge k(\nu+1)\Delta t, \\ 0, & t < k(\nu+1)\Delta t, \end{cases}$$

$$I^{(3)}_n(t,\nu) = D_n(t) - D_n(0).$$

Thus, the expressions for coefficients $\bar{w}_n(t)$ will take the form (4.24):

$$\bar{w}_n(t) = \frac{1}{a_1\beta_n}\bar{I}_n(t) + e^{-\gamma t}\left[\bar{w}_n^{(0)}\cos(\beta_n t) + \frac{1}{\beta_n}\left(\bar{w}_n^{(1)} + \gamma\bar{w}_n^{(0)}\right)\sin(\beta_n t)\right], \tag{4.32}$$

where value $\bar{I}_n(t)$ is determined by expressions (4.31). Figs. 18, 19 show the trajectory of the rod tip for the case of a periodic pulse (4.29) with different relative duration parameters 2 and 5, respectively. Pulse duration $\Delta t = 1s$. For relative duration $S = 2$, a thermal pulse exists for the half of the period, and it is absent for the other half (see Fig. 18). The graph shown in Fig. 18 corresponds to five periods of the pulse function (4.29). In Fig. 18 it can be seen that during the first pulse $0 < t < \Delta t$ the oscillation amplitude is significantly less than during the second ($2\Delta t < t < 3\Delta t$ ), third ($4\Delta t < t < 5\Delta t$ ) and all the consequent pulses. The third, fourth and fifth pulses correspond to almost identical amplitudes of oscillations of the rod tip under the action of thermal strain. Due to the low relative duration, by the beginning of the next pulse, the rod has substantially non-zero initial conditions, which leads to an increase in the oscillation amplitude after the end of the pulse. This fact is clearly seen when comparing Fig. 14 with Figs. 18, 19. The beat amplitude in Fig. 18 remains prominent throughout the period.

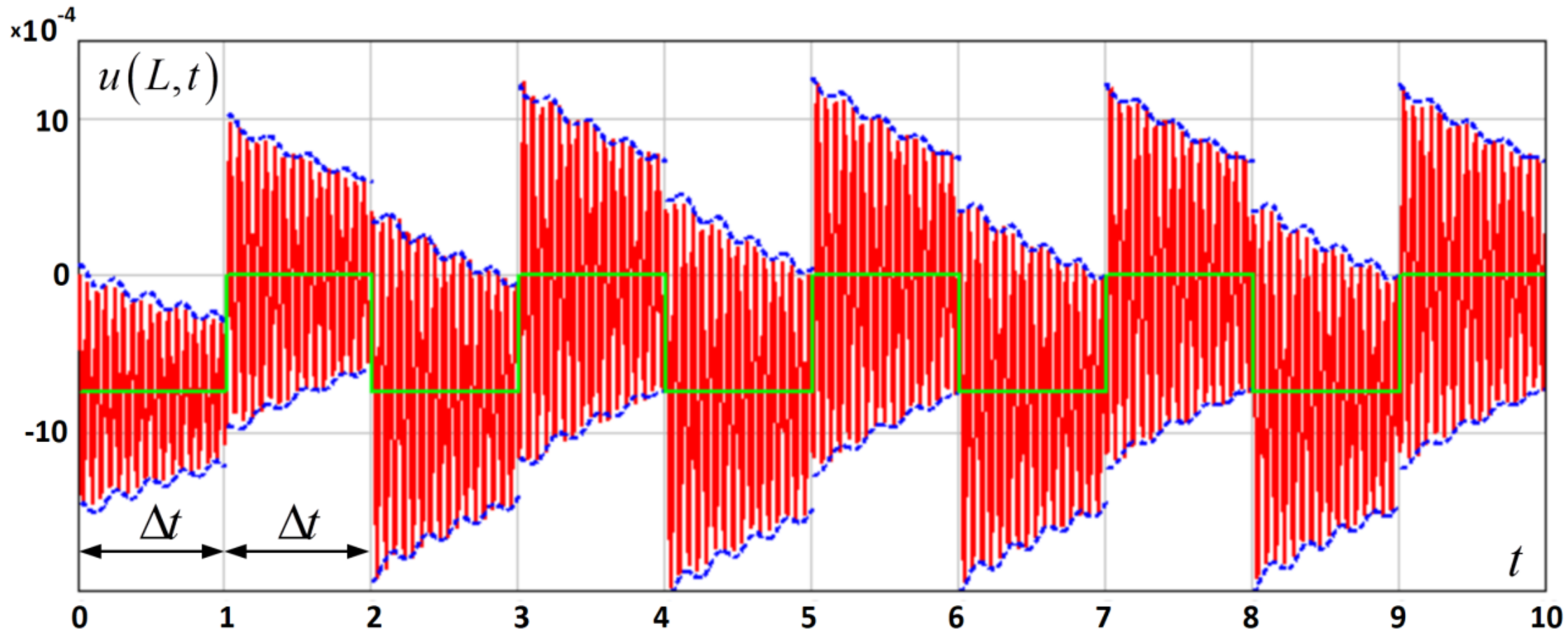

Fig. 18 Trajectory of the rod tip at $S = 2$

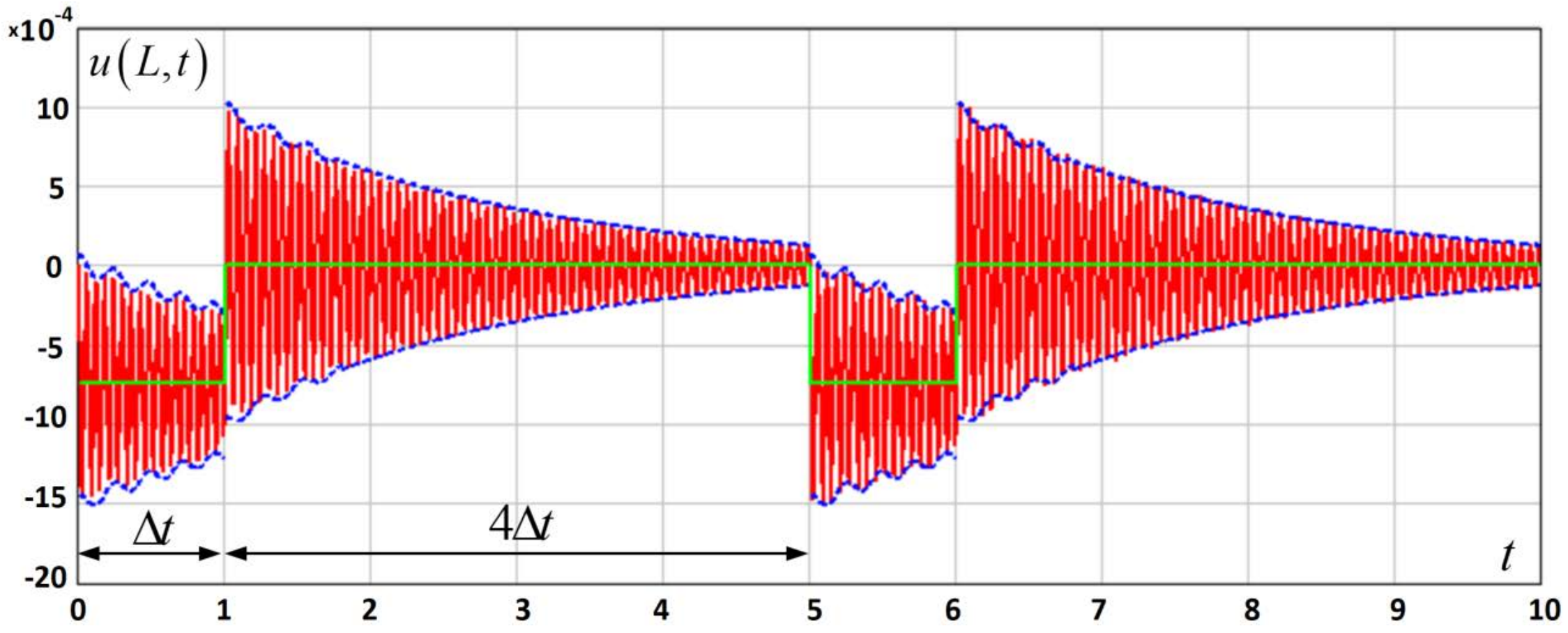

Fig. 19 Trajectory of the rod end at $S = 5$

In Fig. 19 the oscillation graph corresponds to a large relative duration ($S=5$). The two oscillation periods given in Fig. 19 are actually «independent», that is, the behavior of the system at the end of first period $t=5\Delta t$ has little effect on the oscillation process in second period $5\Delta t<t<10\Delta t$. This situation is due to the small values of the initial conditions at the end of the first period.

From the comparisons made in Figs. 18, 19 it can be seen that the correlation of oscillation amplitudes between adjacent pulses increases with a decrease in the relative duration, and decreases with an increase in the relative duration. This fact is expected, since the relative duration actually determines the degree of independence of two neighboring pulses. Indeed, the more time passes between pulses, the more time the system has to reach the initial state of equilibrium as a result of exponential damping.

At the end of the section, let us consider the mathematical reason for beats to occur in Figs. 14-16 and Figs. 18, 19. Without loss of generality, we consider the solution in terms of generalized functions with coefficients (4.32). Under zero initial conditions, the trajectory of the rod end is described by a simple expression:

$$u(L,t)=\frac{2}{a_1}\sum_{n=1}^{+\infty}\frac{(-1)^{n+1}}{\beta_n}\overline{I}_n(t), \tag{4.33}$$

where it is taken into account that $W_n(L)=2(-1)^{n+1}$, value $\overline{I}_n(t)$ is determined by expression (4.31). Fig. 20 shows the absolute values of the first four terms of series (4.33). The dotted line in Fig. 20 shows the monotonic decrease of the amplitude of the function $2\overline{I}_1(t)/a_1\beta_1$. A similar character of behavior is also valid for the remaining values of $n$. Thus, for each summand in the sum (4.33) there are no beats.

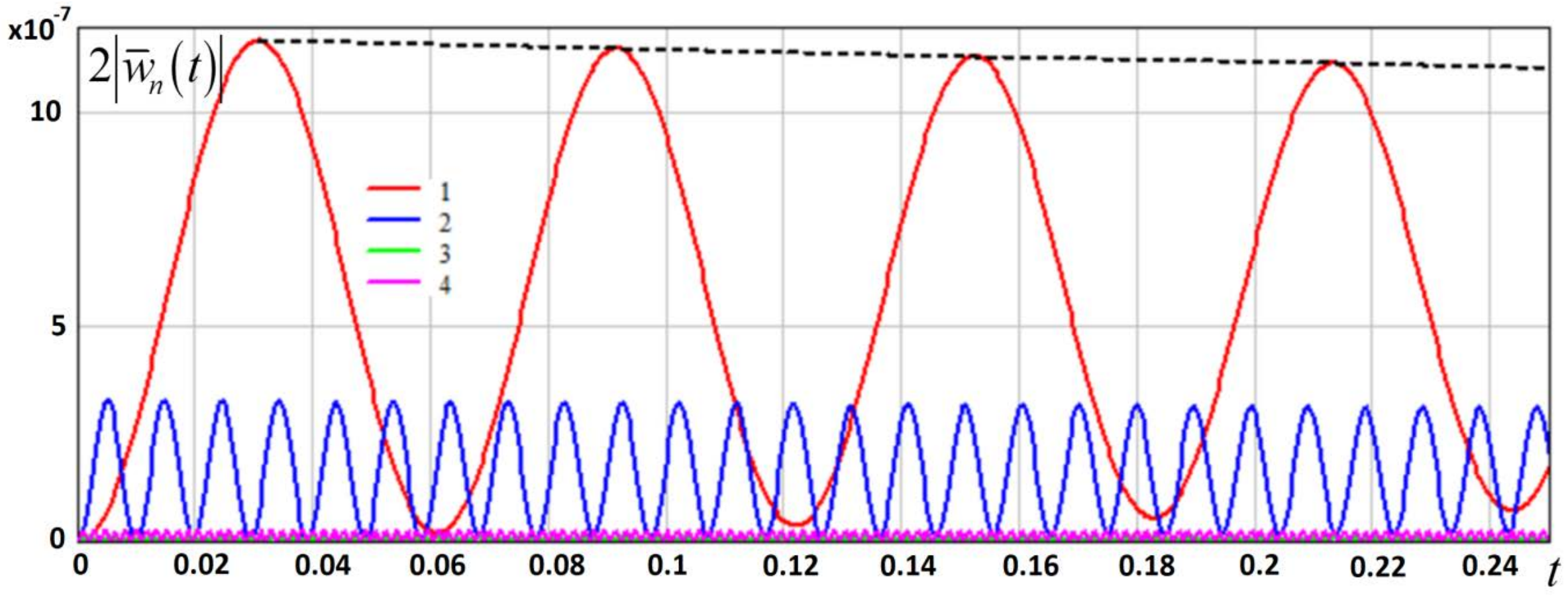

Fig. 20 The first four terms of the series (4.33)

According to expression (4.31), integrals $\overline{I}_n(t)$ are determined in terms of functions $D_n(t)$ that are proportional in order of magnitude to $e^{-\gamma t}/\beta_n$, since in the case under consideration (4.13)-(4.15) $\gamma \ll \beta_n$. The rapid increase in frequencies $\beta_n$ leads to a difference between the amplitudes of functions $2\overline{I}_n(t)/a_1\beta_n$ by almost orders of magnitude (see Fig. 20). Indeed, taking into account (3.16) and estimating $|\overline{\tau}_n''| \sim \alpha_n^2$ (4.27), we obtain:

$$\beta_n \approx \frac{\alpha_n^2}{L^2\sqrt{a_1}} \sim n^2, \quad \left|D_n(t)\right| \approx \frac{L^2\sqrt{a_1}}{\alpha_n^2} e^{-\gamma t}, \quad \left|\overline{I}_n(t)\right| \sim \frac{\overline{\alpha}_T\left|\overline{\tau}_n''\right|}{d_0 L^2}\left|D_n(t)\right| \sim \frac{\overline{\alpha}_T}{d_0}\sqrt{a_1} e^{-\gamma t},$$

$$\frac{2}{a_1\beta_n}\left|\overline{I}_n(t)\right| \sim \frac{e^{-\gamma t}}{n^2}. \tag{4.34}$$

The majorizing estimate (4.34) implies the absolute convergence of series (4.33) over the entire time interval. The rapid convergence of series (4.33) is valuable in numerical calculations and makes it possible to take into account, for example, only the first two terms.

The result of such a summation for solution (4.33) is shown in Fig. 21. As can be seen in Fig. 21 the superposition of the first two harmonics leads to the behavior characteristic of the beat process. Since the amplitudes of the harmonics (4.34) and their frequencies $\beta_1$ and $\beta_2$ differ by almost four times, then in Fig. 21, the classical pattern of beats cannot be seen, but a non-monotonic change in the envelope of the oscillation amplitude is observed. Accounting for the remaining terms of series (4.33) will not fundamentally change the form of the trajectory shown in Fig. 21, since, according to estimate (4.34), their contribution will be of the order of $\sim 1/n^2$.

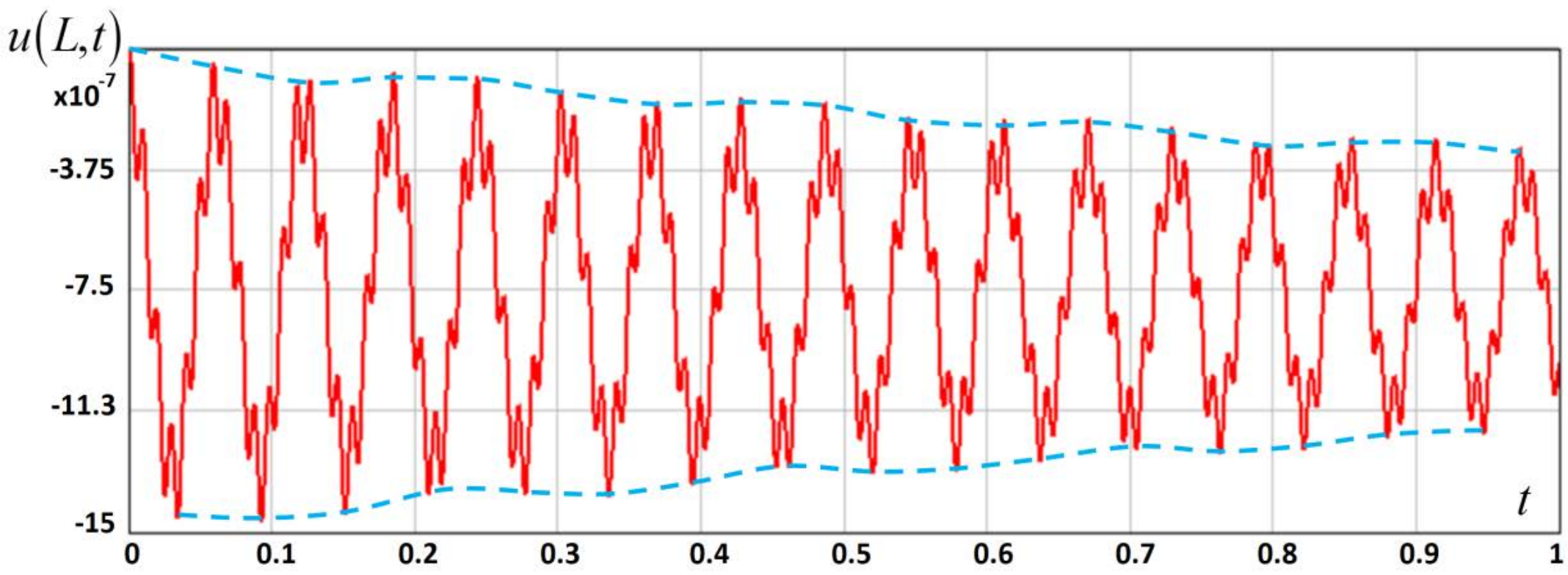

Fig. 21 Sum of the first two terms of series (4.33)

In fact, it turns out that beats are present during the entire time of oscillations, but at large times, due to exponential damping (4.34), their contribution becomes invisible in Figs. 14-16 and Figs. 18, 19.

No special numerical methods were necessary to calculate our analytical solution. All the calculations were performed by C++ programming language with the use of MathCAD package software.

## §5 Discussion

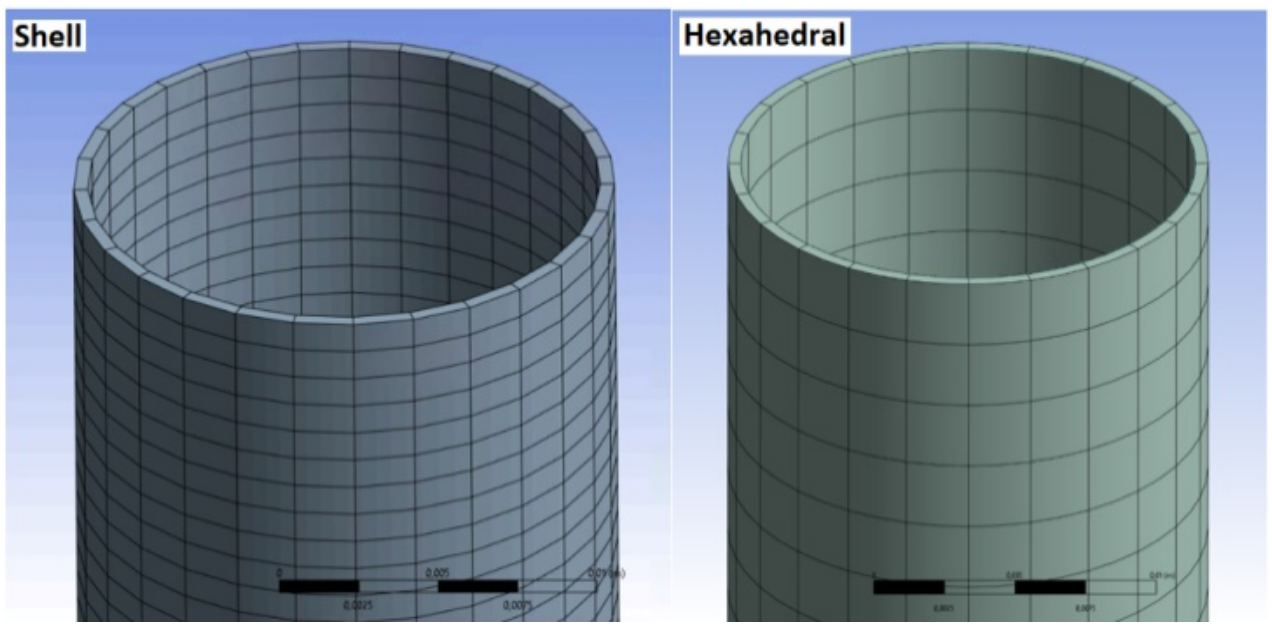

Fig. 22 Mesh element type

Let us perform a comparative analysis of the obtained analytical solution and the numerical one made with the ANSYS software. Fig. 22 shows two types of finite-element mesh in ANSYS: 10 000 shell elements (on the left) and 30 000 hexahedral elements (on the right). The basis functions of the finite elements are of the second order of approximation. The cylinder (see Figs. 1, 22) is thin-walled, its thickness is equal to the thickness of the element.

The heat exchange between rod-shell and the environment as well as inside the rod itself is neglected when calculating deformations. The initial rod temperature is taken equal to $T_0 = const$ all over its volume. Its upper part is rigid fastened and weight force is neglected. Viscous resistance force is not taken into account. Material lengthening is considered to be linear dependent on the temperature (1.13). The pulses' duty cycle $S = 1$ is used to compare ANSYS results with the analytical calculation data. We set initial conditions for each time step. The FE mesh is divided by three parts. The central part is $L_c = 0.4m$ long, the rest parts (symmetrical to the central part) are $L_{1,2} = 0.3m$ long (see Fig. 3). At the heat pulse moment of time the temperature of the central part (mesh nodes) is set according to (1.7) $T = \overline{T} - T_0 = \frac{2\Delta T}{d_0} y$, where $\Delta T = 0.5K$. The temperature of the rest parts is set to $\overline{T} = T_0$. Between the heat pulse moments, the whole rod temperature is taken equal to $\overline{T} = T_0$.

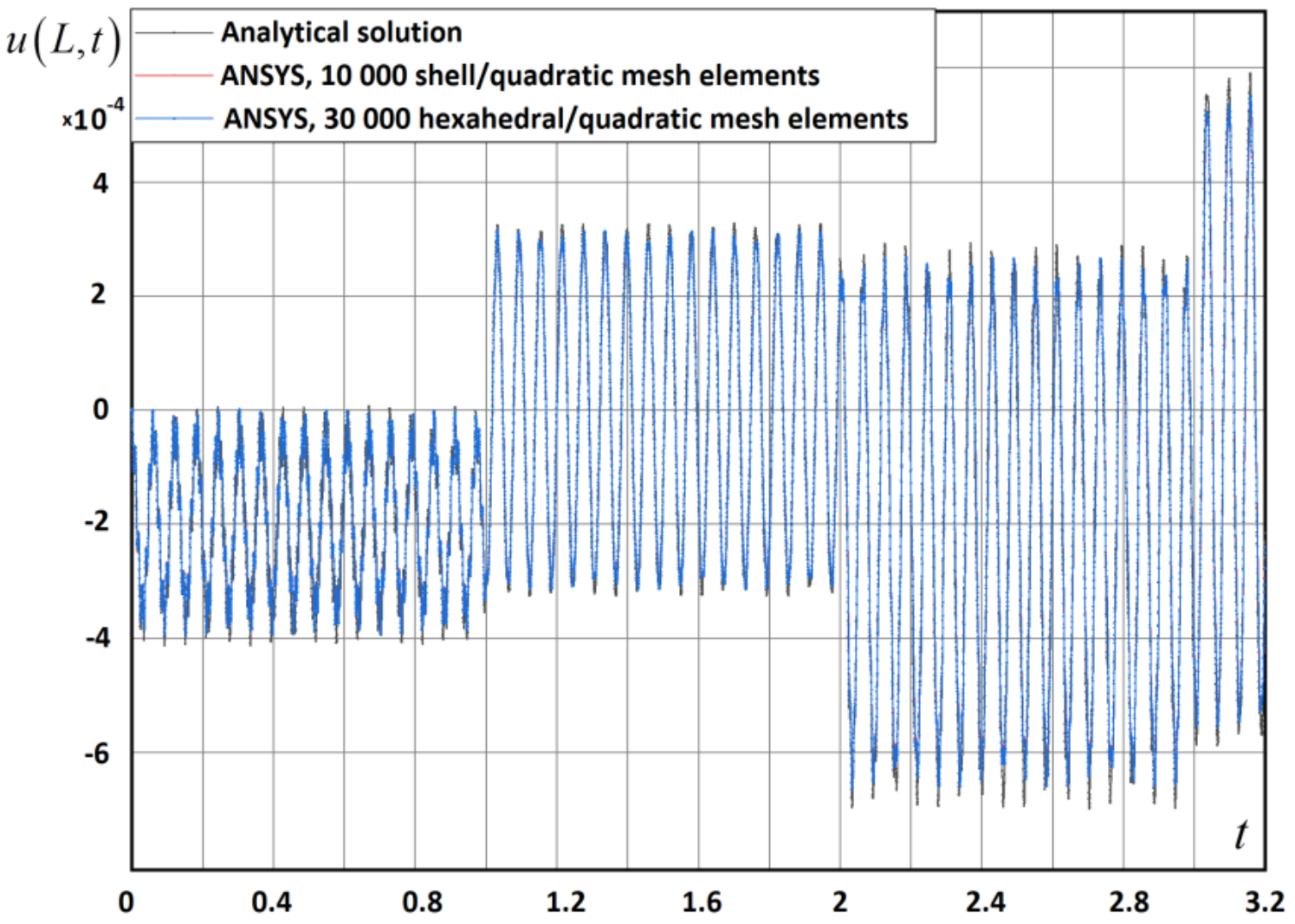

Fig. 23 Comparison of ANSYS results with analytical solution

As shown by the analytical calculations (4.33)-(4.34) (see Fig. 20) the dominating form of the rod tip oscillation is sine-shaped with a period of about $t = 0.06s$, but the «period» of the sine curve peaks is estimated to be $t = 0.008s$. Thus, we constrained the time step for the numerical calculations by ANSYS. The maximum step is set 20 times lower to be equal to $dt_{max} = 0.0004s$. The stepping was taken automatically by ANSYS within the limits from $dt = 0.0001s$ to $dt_{max} = 0.0004s$. An initial step of $dt = 10^{-5}s$ was taken within $\pm 0.05s$ of time vicinity of pulses to get the first oscillation curve with a higher accuracy. Fig. 23 shows rod tip oscillation curves for three cases: one analytical and two numerical ANSYS (Shell/Hexahedral) calculations. Fig. 23 shows the behavior of the rod tips for the time period of $3.2s$. This time period contains two heat pulses. All the three curves on Fig. 23 seem to be almost the same.

Hence, the analytical solution obtained is close to the direct numerical ANSYS calculation with a high accuracy. An increase in the oscillation amplitude with each next heat pulse can be clearly seen on Fig. 23. The reason for this behavior is the resonance effect. As a result the amplitude of oscillation increases with time.

Table 1

| **Element type** | **Shell/quadratic** | **Hexahedral/quadratic** |
|---|---|---|
| Number of elements | 10 000 | 30 000 |
| Number of steps | 8 171 | 8 133 |
| Calculation time | 2 h. 20 min. | 7 h. 15 min. |

To reach a reasonably enough accuracy of the reactor simulation it is necessary to calculate more than 50 fuel elements [20-21]. Table 1 shows that the numerical solution of the thermoelastic problem is time-consuming. Analytical calculation is much faster and takes less than one minute. Thus, the analytical solution presented in this paper appears to be rather efficient for the reactor stability analysis.

## Conclusion

Currently, it is impossible to run the IBR-2 reactor at its maximum power due to the instability induced by the FE dynamic bending effect. The oscillations make the FE to move further from or closer to the reactor core which affects the setup reactivity. We propose to solve this problem both for this reactor and future next-generation NEPTUNE reactor, which project now is underway, by taking the FE dynamic bending effect into account when calculating the reactivity and other parameters, providing the stable operation. To reach a reasonable accuracy it is necessary to simulate a system with more than 50 FE which is a time-consuming and resource-intensive problem. And this problem can be solved with the analytical solution presented in this paper which is several orders faster as compared with usually widely used numerical analysis such as direct numerical calculations by ANSYS (see Discussion).

Our results can open the possibility to approach even a more complicated problem, namely an estimation of mutual influence of fuel elements on the reactor stability. For example, our fuel element dynamics model can be extended by adding the effect of a vibrating mount. In a particular case, such a problem is reduced to the known Kapitsa pendulum [14-15] or a set of its generalizations [16-19].

## Acknowledgments

The authors are grateful to Corresponding Member of the Russian Academy of Sciences V.L. Aksenov for a fruitful discussion of the physical aspects of the reactor theory and a number of valuable additions to the material of the paper.

This research has been supported by the Interdisciplinary Scientific and Educational School of Moscow University «Photonic and Quantum Technologies. Digital Medicine».

The authors deeply acknowledge the referees of this paper for their valuable comments and proposals.